\documentclass[fleqn,usenatbib]{mnras}

\usepackage{newtxtext,newtxmath}

\usepackage[T1]{fontenc}
\usepackage{ae,aecompl}

\usepackage{graphicx}	
\usepackage{amsmath}	
\usepackage{amssymb}	
\usepackage{times}

\title[The Edge of the Galaxy]{The Edge of the Galaxy}

\author[A. J. Deason et al.]{
Alis J. Deason$^{1}$\thanks{E-mail: alis.j.deason@durham.ac.uk}, Azadeh Fattahi$^{1}$, Carlos S. Frenk$^{1}$, Robert J. J. Grand$^{2}$, \newauthor Kyle A. Oman$^{1}$, Shea Garrison-Kimmel$^{3}$, Christine M. Simpson$^{4,5}$, Julio F. Navarro$^{6}$
\\
$^{1}$Institute for Computational Cosmology, Department of Physics, University of Durham, South Road, Durham DH1 3LE, UK\\
$^{2}$Max-Planck-Institut f{\"u}r Astrophysik, Karl-Schwarzschild-Str. 1, D-85748, Garching, Germany\\
$^{3}$Factual Inc., 1999 Ave of the Stars, Los Angeles, CA 90067\\
$^{4}$Enrico Fermi Institute, The University of Chicago, Chicago, IL 60637, USA\\
$^{5}$Department of Astronomy and Astrophysics, and Kavli Institute for Cosmological Physics, The University of Chicago, Chicago, IL 60637, USA\\
$^{6}$Department of Physics and Astronomy,University of Victoria, PO Box 3055 STN CSC, Victoria, BC, V8W 3P6, Canada
}

\date{Accepted XXX. Received YYY; in original form ZZZ}

\pubyear{2020}

\begin{document}
\label{firstpage}
\pagerange{\pageref{firstpage}--\pageref{lastpage}}
\maketitle

\begin{abstract}
  We use cosmological simulations of isolated Milky Way-mass galaxies, as well as Local Group analogues, to define
  the ``edge'' --- a caustic manifested in a drop in density or
  radial velocity --- of Galactic-sized haloes, both in dark matter
  and in 
  stars. In the dark matter, we typically identify two caustics: the
  outermost caustic located at $\sim \! \! 1.4r_{\rm 200m}$
   corresponding to the ``splashback'' radius,
  and a second caustic located at $\sim \! \! 0.6r_{\rm 200m}$
   which likely corresponds to the edge of the
  virialized material which has completed at least two pericentric passages. The splashback radius is ill defined in Local Group
  type environments where the halos of the two galaxies overlap.
  However, the second caustic is less affected by the presence of a
  companion, and is a more useful definition for the boundary of
  the Milky Way halo. Curiously, the stellar distribution also has a
  clearly defined caustic, which, in most cases, coincides with the
  second caustic of the dark matter. This can be identified in both
  radial density and radial velocity profiles, and should be
  measurable in future observational programmes. Finally, we show that
  the second caustic can also be identified in the phase-space
  distribution of dwarf galaxies in the Local Group. Using the current
  dwarf galaxy population, we predict the edge of the Milky Way halo to be
  $292 \pm 61$ kpc.
\end{abstract}


\section{Introduction}

The mass condensations commonly referred to as dark matter haloes in simulations fade
gradually into the background matter distribution and have no
well-defined edge \citep[e.g.][]{diemer13}. Furthermore, haloes are not spherical but have
irregular shapes. Nevertheless, definitions of the nominal boundary of
a halo such as the ``friends-of-friends'' radius \citep{davis85}, the
``virial radius'' \citep[e.g.][]{cole96} or ``$r_{200}$'' abound in the
literature. Even the latter is ambiguous, as it is sometimes defined
as the radius, $r_{\rm 200c}$, within which the mean density equals
200 times the critical density \citep[e.g][]{nfw96} or as the radius,
$r_{\rm 200m}$, within which the mean density equals 200 times the mean
cosmic value \citep[e.g.][]{diemand07}.

From a practical point of view, the ambiguity regarding the definition
of the boundary of a dark matter halo can become troublesome when we
want to define the dark matter particles, stars, gas or subhaloes that
``belong'' to a halo, or when we wish to define the radius at which
tracers can escape from a self-bound system \citep[e.g.][]{lt90,springel05}. The
physical extent of haloes varies significantly at different mass
scales and in different environments \citep[e.g.][]{nfw96, nfw97,
  bullock01, wechsler02} and, when contrasting simulations or
comparing them to observations, a common definition of halo extent is
essential to avoid confusion.  In addition, while the backdrop of our
current theory of structure formation is cold dark matter, it is just
as important to understand how the baryonic components relate to the
dark matter, and where observational boundaries lie
\citep[e.g.][]{kravtsov13, shull14, wechsler18}.

Analytical solutions for the collapse of spherical gravitational
structures in a cosmological context provide valuable insight into the
structure of dark matter haloes. The spherical collapse model, first
presented by \cite{gunn72} for an Einstein-de Sitter Universe,
describes the evolution of spherical shells of matter around an
overdensity \citep[see also][]{fillmore84, bertschinger85}. In this
model, initially overdense regions gravitationally attract the
surrounding matter, causing it to detach from the Hubble flow and
collapse, forming larger and larger equilibrium structures. Each
successive mass shell collapses onto a deeper potential well and thus
has a higher energy and a larger apocentre. Material piles up at these
apocentres, giving rise to a singularity or caustic surface. Of
particular interest is the outermost caustic, termed the
``splashback'' radius, which corresponds to the apocentre of  material that has most recently completed its first pericentric passage.

The spherical collapse model has served as a motivation for many of
the commonly used definitions of halo masses and sizes. Traditionally
(see e.g. \citealt{binney08} Section 9.2.1), an Einstein-de Sitter
Universe is assumed, where energy conservation and the virial theorem
imply that the ``virial'' radius (enclosing the mass whose potential
energy is twice the negative kinetic energy) occurs at half the
turnaround radius. In the Einstein-de Sitter model the overdensity
(relative to the critical density) at virialization is
$\Delta_{\rm c} = \rho_{\rm vir}/\rho_{\rm c} =18 \pi^2 = 178$. This
formalism has been generalized for a $\Lambda$CDM universe
\citep{lahav91, eke98, bryan98}, in which case the overdensity defining the
boundary is $\Delta_{\rm c} \sim 100$ at $z=0$, and evolves with
redshift.

In the spherical collapse model the virial radius defines the region
within which the system is virialized; beyond this radius mass is
still collapsing onto the object. N-body simulations suggest that this
distinction occurs at $\Delta_{\rm c} \sim 200$ \citep{cole96}, so a
commonly used definition of halo is $r_{\rm 200c}$.  Another commonly
used definition, particularly in studies of the halo occupation
distribution of galaxies \citep[e.g.][]{berlind02, kravtsov04}, is
$r_{\rm 200m}$, which corresponds to
$\Delta_{\rm c}=200 \times \Omega_{\rm m} \sim 60$ today. For a Milky
Way mass halo ($\sim \! \! 1 \times 10^{12} \mathrm{M}_\odot$), these halo boundaries
are typically: $r_{\rm 200c} \approx 220$ kpc,
$r_{\rm vir} \approx 290$ kpc, and $r_{\rm 200m} \approx 350$
kpc. Several authors have argued that the splashback radius, predicted
by the spherical collapse model, is the most natural definition of the
boundary of a halo \citep[e.g][]{adhikari14, diemer14, more15}. For a
Milky Way halo the splashback radius is typically $\sim \! \! 500$ kpc (assuming the splashback radius lies at $\sim \! \! 1.5 r_{\rm 200m}$, see below).

In reality, halo collapse is non-spherical, lumpy and significantly
anisotropic. Several works have used N-body simulations to follow this
collapse in detail \citep[e.g.][]{davis85, frenk88,cole96,diemand08,
  springel08} and to compare with the predictions of the spherical
collapse model \citep[e.g.][]{prada06,zavala08,ascasibar07,ludlow10}.
While most studies have concentrated on the inner profiles of dark
matter haloes \citep[e.g.][]{nfw96, moore99, stadel09}, more recently,
\cite{adhikari14}, \cite{diemer14} and \cite{more15} have explored the outer density
profiles of dark matter haloes. These studies identify the outer
caustic, or splashback radius, as a sharp jump in the density
profile. For example, \cite{diemer14} and \cite{more15} find that the splashback radius
falls in the range $(0.8-1.0)r_{200 \rm m}$ for rapidly accreting
haloes, and is $\approx 1.5r_{200 \rm m}$ for slowly accreting
haloes. 

The influence of environment, mass accretion rate, and redshift on the
splashback radius was investigated by \cite{diemer17} and
\cite{mansfield17} and the splashback radius is now a commonly used,
and thoroughly explored halo boundary. Interestingly, there is now
considerable evidence that splashback radii have been measured
observationally in the outskirts of galaxy clusters
\citep[e.g][]{more16, baxter17, chang18, shin19, contigiani19, zurcher19, murata20}. While the measured
splashback radii tend to be smaller than those predicted in
$\Lambda$CDM simulations, these results are still subject to
systematic effects \citep{busch17, xhakaj19, murata20}.

Often the most relevant, and even the most physical, definition of
halo boundary depends on the situation at hand. The term splashback is
often used by reference to the population of ``backsplash'' galaxies,
i.e. galaxies that have been inside, but are now outside the virial
radius, and may extend well beyond any traditional spherical collapse boundary \citep[e.g.][]{balogh00, mamon04, gill05, sales07, ludlow09, teyssier12, bahe13, wetzel14}.
The properties of these backsplash galaxies demonstrate that the
environmental effects of haloes can extend well beyond the traditional
virial radius boundary. However, even if the zone of influence of
haloes extends significantly beyond the virial radius, haloes are
never isolated systems, and eventually run into other massive
systems. For example, the Milky Way galaxy resides in the Local Group,
and is located $\sim \! \! 800$ kpc from the roughly equal mass halo of
M31. Thus, the splashback radius for a Milky Way mass halo runs into
that of M31. In this case, it is perhaps more physical to consider the
splashback radius of the entire Local Group, rather than of its
individual components. Nonetheless, a physically motivated definition
of the extent for the Milky Way is warranted, and will become even
more important when the next generation surveys discover many tens of
dwarf galaxies in the Local Group.
 
In this work we explore the boundary of Milky Way mass haloes using
high-resolution cosmological simulations. In particular, we use the
outer density profiles of the haloes to quantify their extent. We take
into account two important characteristics of the Milky Way: (1)~its
location in the Local Group, and hence its proximity to M31, and
(2)~the relation between the extent of the stellar distribution and
that of the underlying dark matter. This consideration is important
for observational probes of the Milky Way halo boundary. In
Section~\ref{sec:sims} we describe the cosmological simulations used
in this work. These comprise both collisionless and hydrodynamic
simulations, as well as simulations designed to mimic the Local
Group. We quantify the ``edges'' of the dark matter haloes, stellar
haloes, and satellite dwarf galaxy populations, and compare these
various boundaries in Section~\ref{sec:edges}. Finally, we summarise
our main results in Section~\ref{sec:conc}.
 
 \section{Simulations}
 \label{sec:sims}
We use a large range of high resolution simulations of Milky Way-mass
haloes to quantify the edges of Galactic-sized haloes. Below we
describe each simulation suite in turn.  

\subsection{ELVIS}
The ``Exploring the Local Volume in Simulations'' (ELVIS) project is a
suite of 48 simulations of Galaxy-size haloes \citep{gk14}. These
simulations were designed to model the Local Group (LG) environment in
a cosmological context. Half of the haloes (24) are in paired
configurations similar to the Milky Way and M31. The LG analogues were
selected from medium resolution
($m_p=9.7 \times 10^7 \mathrm{M}_\odot$, force softening 1.4 kpc)
cosmological simulations. Twelve halo pairs were selected for
resimulation based on phase-space criteria appropriate to the MW/M31
system (e.g. separation, total mass, radial velocity). The resulting
zoom simulations are high resolution
($m_p = 1.9 \times 10^5\mathrm{M}_\odot$, force softening 141 pc)
volumes that span 2-5 Mpc in size. The remaining half (24) of the
ELVIS suite are isolated, mass-matched analogues, which are
resimulated at the same resolution as the paired haloes. The resulting
sample consists of 48 high-resolution haloes in the mass range
$1-3 \times 10^{12}\mathrm{M}_\odot$. The ELVIS suite was run with the
\textit{WMAP-7} cosmology \citep{larson11} with parameters:
$\Omega_{\rm M} = 0.266$, $\Omega_{\Lambda} =0.734$, $H_0 =71$ km
s$^{-1}$ Mpc$^{-1}$.

Subhaloes were identified using the \textsc{rockstar} halo finder \citep{behroozi13a} and were followed through time with \textsc{consistent trees} \citep{behroozi13b}. We define the centre of the host haloes using the position and velocity of the main subhalo calculated in the \textsc{rockstar} algorithm. \cite{gk14} find that the subhalo sample in ELVIS is complete down to $M_{\rm sub} > 2 \times 10^7 \mathrm{M}_\odot$ (or $V_{\rm max} > 8$ km s$^{-1}$). The general properties of the ELVIS haloes are described in \cite{gk14} and summarised in their Table 1. This suite has produced a number of results, including predictions for future dwarf galaxy detections \citep{gk14}, the stellar-mass halo relation for LG galaxies \citep{gk17}, the prevalence of dwarf-dwarf mergers and group-infall onto MW mass haloes \citep{deason14b, wetzel15}, and insights into the planar alignment of MW satellites \citep{pawlowski17}.

\subsection{APOSTLE}
APOSTLE (A Project Of Simulating The Local Environment) is a suite of
high resolution, hydrodynamic simulations consisting of 12 halo pairs
\citep{fattahi16, sawala16}. These pairs were drawn from the medium
resolution ($m_p = 8.8 \times 10^6\mathrm{M}_\odot$) \textsc{dove}
dark matter-only cosmological simulation described by \cite{jenkins13}.  The candidates were selected to have paired configurations similar to the LG, based on the separation of the pairs, their relative radial and tangential velocities, a Hubble flow constraint, and the combined mass of the pair. The exact selection criteria differ from the ELVIS suite, with the main difference being the total masses of the haloes. The APOSTLE suite has typically lower halo masses, and span the mass range $0.5-2.5 \times 10^{12}\mathrm{M}_\odot$. The resimulations span 2-3 Mpc in size and were run with the same hydrodynamic code as the EAGLE Reference calibration \citep{schaye15, crain15}, which includes subgrid prescriptions for star formation, feedback, metal enrichment, cosmic reionization, and AGN. The simulations were performed at three different resolution levels, and we use the ``medium'' L2 resolution suite which has 10 times better mass resolution than \textsc{dove} ($m_p = 6 \times 10^5\mathrm{M}_\odot$, force softening 307 pc), with a gas particle mass of $1.2 \times 10^5\mathrm{M}_\odot$. APOSTLE was run with the \textit{WMAP-7} cosmology  \citep{komatsu11} with parameters: $\Omega_{\rm M} = 0.272$, $ \Omega_{\rm b} = 0.0455$, $\Omega_{\Lambda} =0.728$, $H_0 =70.4$ km s$^{-1}$ Mpc$^{-1}$.

Haloes are identified using a friends-of-friends (FOF) algorithm \citep{davis85}, and subhaloes belonging to each FOF halo were identified using the \textsc{subfind} algorithm \citep{springel01}. We use the position and velocity of the main (sub)halo calculated in \textsc{subfind} to define the centre of the host halo. Note that this definition of halo centre is different to the one used in ELVIS, which is based on \textsc{rockstar}. For a comparison of the \textsc{subfind} and \textsc{rockstar} subhalo finding algorithms, see e.g. \cite{knebe11}.
\cite{sawala16} showed that the satellite luminosity function of APOSTLE L2 is complete down to $M_{\rm star} \sim 10^5\mathrm{M}_\odot$, and they used the APOSTLE suite to address apparent small-scale problems in the $\Lambda$CDM cosmology. In particular, they showed that the simulations match the abundance of observed dwarf satellites in the Milky Way and M31, thus solving the apparent ``missing satellites'' \citep{moore99a} and ``too-big-to-fail'' \citep{bk11} problems. Several other works have used the APOSTLE suite to investigate a wide range of topics. These include, probing the nature and properties of dark matter \citep{lovell17, sawala17},  the tidal stripping of dwarf galaxies and formation of the stellar halo \citep{starkenburg17, fattahi18},  and tests of observational mass estimates of dwarf galaxies \citep{campbell17, genina18,genina19}.

\subsection{Auriga}
The Auriga suite consists of cosmological hydrodynamical zoom-in
simulations of isolated Milky Way-mass haloes
\citep{grand17}. Candidates for resimulation were selected from the
100 cMpc dark matter only cube  of the EAGLE simulation
\citep{schaye15}. The sample of Auriga haloes was chosen to be
relatively isolated at $z=0$,  with no objects with masses greater
than half of the parent halo closer than 1.37 Mpc. The initial sample
of 30 haloes was selected in the mass range $1-2 \times 10^{12}
\mathrm{M}_\odot$, and a further 10 lower mass ($0.5-1 \times
10^{12}\mathrm{M}_\odot$) haloes were more recently added to the suite
\citep{grand19b}. The zoom resimulations were performed with the
\textsc{arepo} code, which follows magnetohydrodynamic and
collisionless components in a cosmological context. At the resolution
used in this work (L4) the gravitational softening is $370$ pc and the
typical particle/cell masses are $3 \times 10^{5}\mathrm{M}_\odot$ and
$5 \times 10^{4}\mathrm{M}_\odot$ for the dark matter and gas,
respectively. The Auriga galaxy formation model includes subgrid
prescriptions for several important physical processes, such as star
formation,  supernova feedback, gas cooling, metal enrichment and
magnetic fields (see \citealt{grand17} for more details). The Auriga
suite was run with the \textit{Planck} cosmology \cite{planck14} with
parameters: $\Omega_{\rm M} = 0.307$, $ \Omega_{\rm b} = 0.048$,
$\Omega_{\Lambda} =0.693$, $H_0 =67.77$ km s$^{-1}$. 

Subhaloes in the Auriga haloes are identified using the \textsc{subfind} algorithm, and we use the position and velocity of the main subhalo calculated by \textsc{subfind} to define the centre of the host.
The Auriga galaxies match well a number of observed properties of disc galaxies, such as their sizes, rotation curves, stellar masses, chemistry and star formation rates \citep{grand16, grand17, marinacci17, grand18}. In addition, the suite has been used to study the stellar haloes of disc galaxies \citep{monachesi16, monachesi19},  interpret the assembly history of the Milky Way halo \citep{deason17, fattahi19a, belokurov20}, study the quenching of satellite galaxies \citep{simpson18}, and measure the total mass of the Galaxy \citep{deason19, grand19, callingham19}.

\section{The Edge of Milky Mass Haloes}
\label{sec:edges}
\begin{figure*}
    \begin{minipage}{0.49\linewidth}
        \centering
        \includegraphics[width=\textwidth,angle=0]{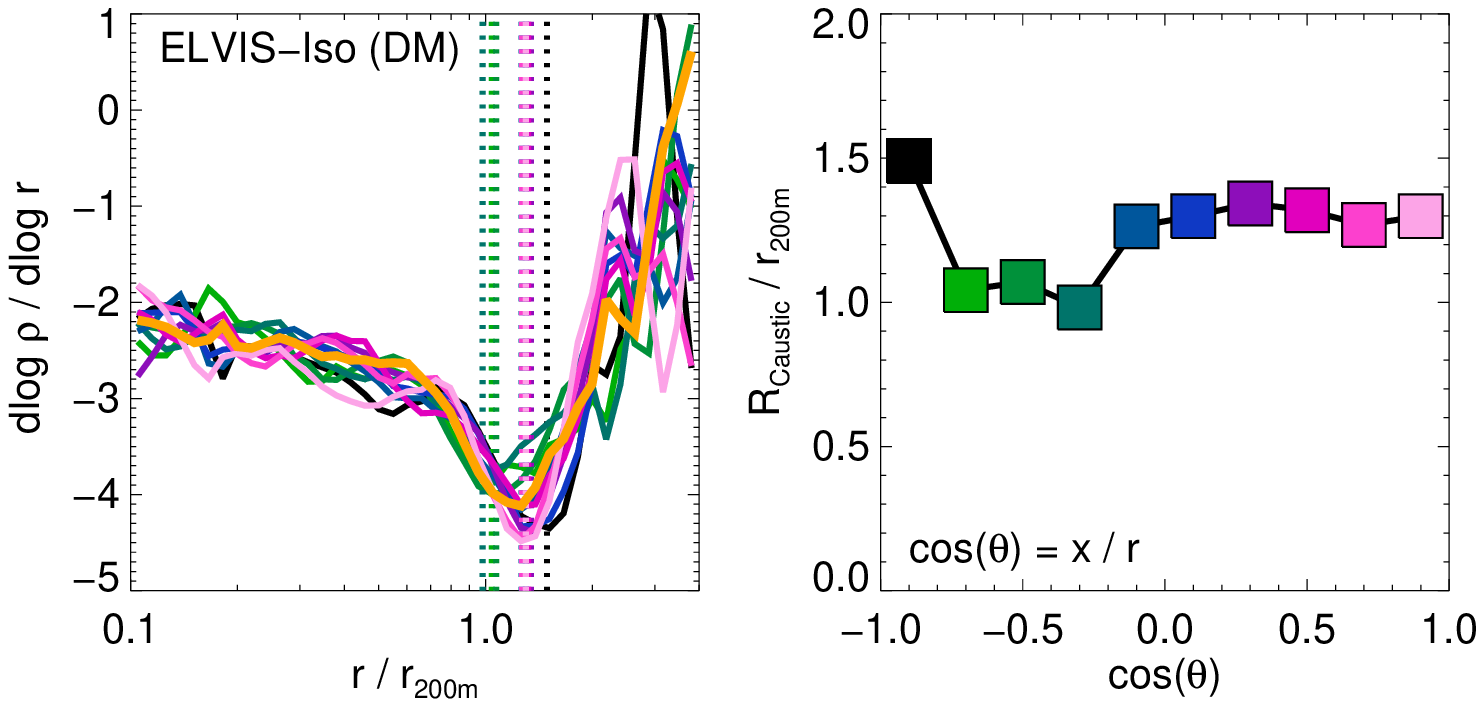}
    \end{minipage}
     \begin{minipage}{0.49\linewidth}
        \centering
        \includegraphics[width=\textwidth,angle=0]{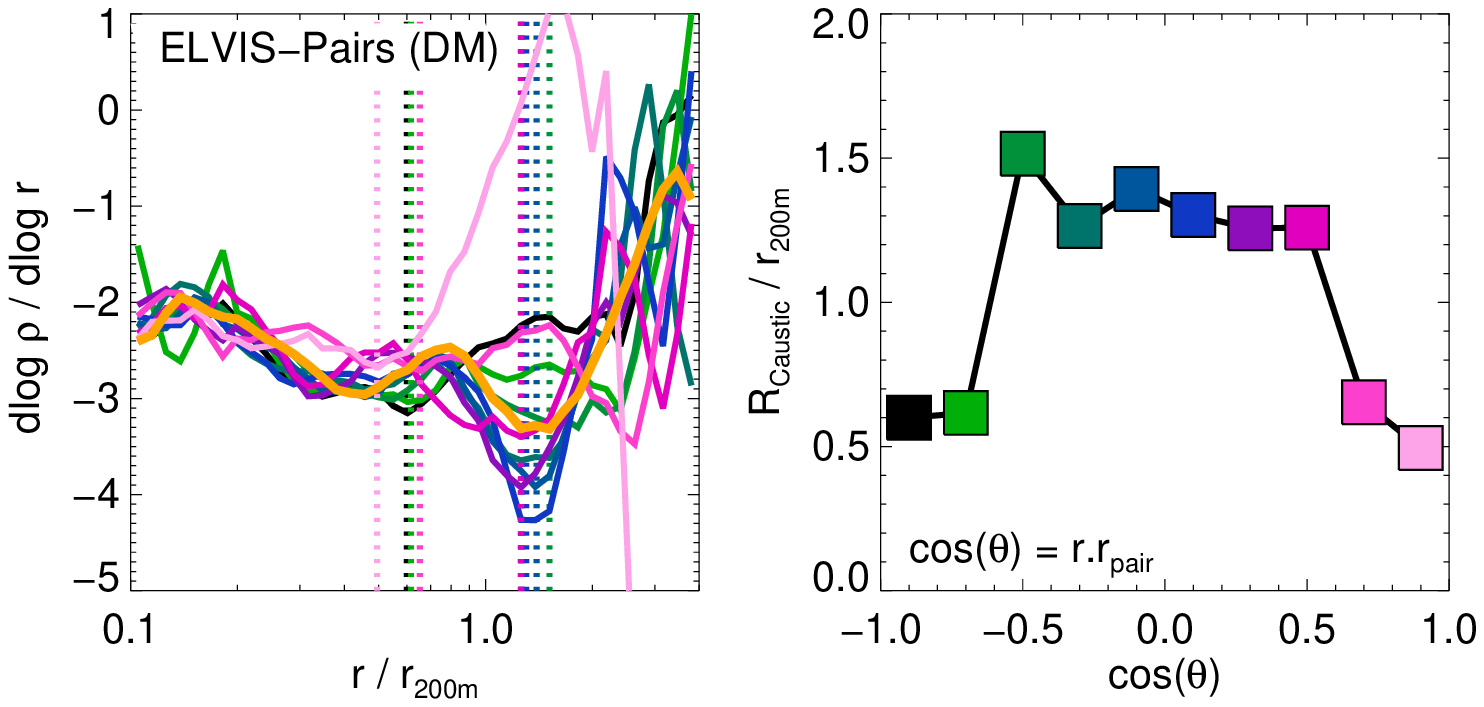}
    \end{minipage}
    \begin{minipage}{0.49\linewidth}
        \centering
        \includegraphics[width=\textwidth,angle=0]{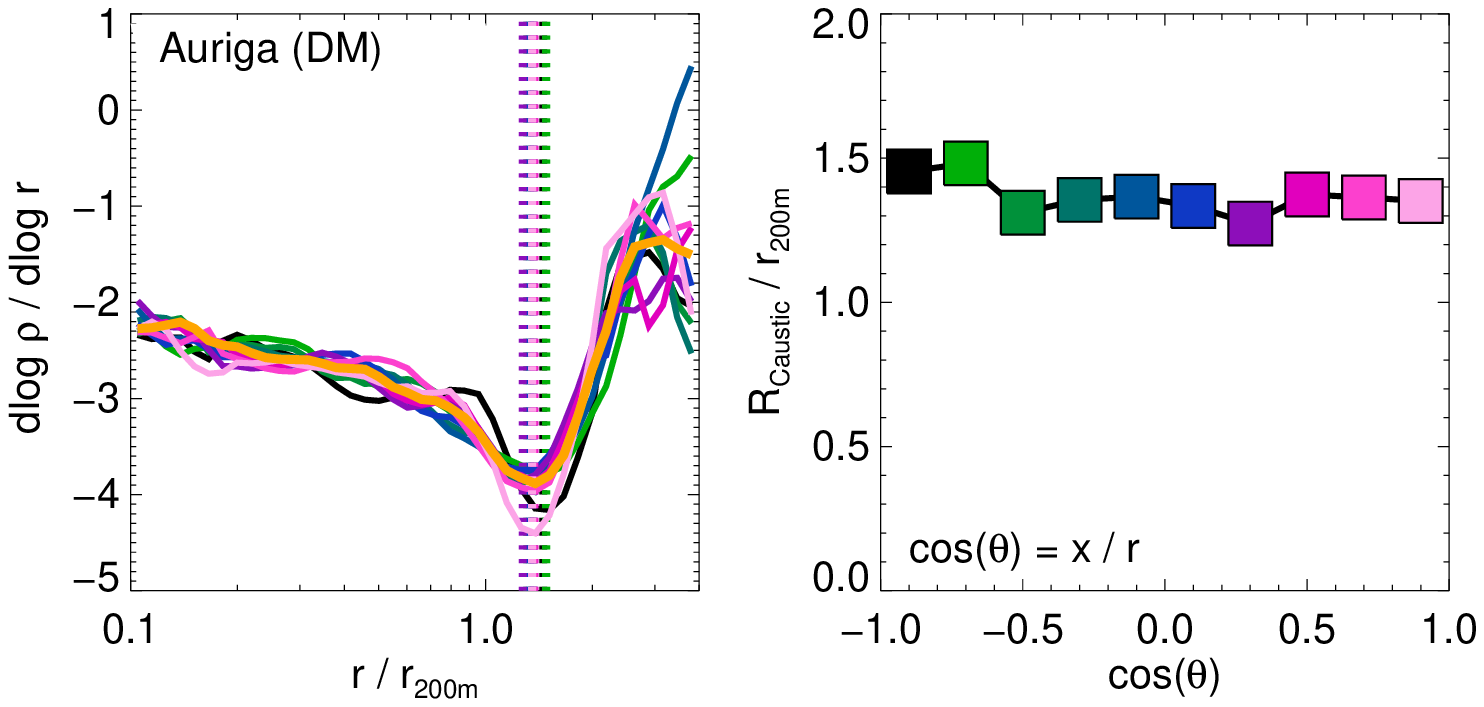}
    \end{minipage}
    \begin{minipage}{0.49\linewidth}
        \centering
        \includegraphics[width=\textwidth,angle=0]{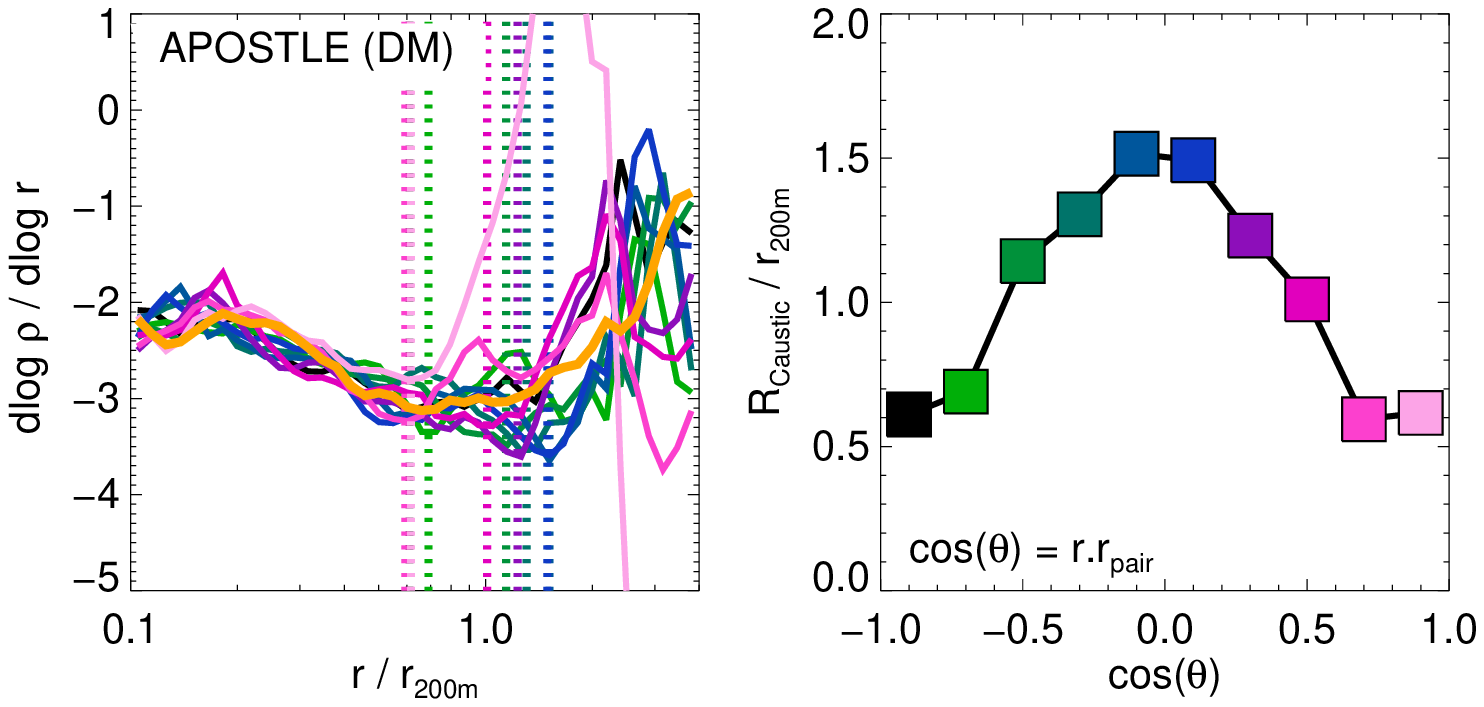}
    \end{minipage}
    \caption[]{The logarithmic slope profile,
      $\mathrm{d\,log}(\rho)/\mathrm{d\,log}(r)$, of the dark matter
      density profiles for the ELVIS (isolated \& paired), Auriga and
      APOSTLE haloes. Here, 40 evenly spaced bins in
      $\mathrm{log}(r/r_{\rm 200m})$ have been used in the range
      $\mathrm{log}(r/r_{\rm 200m}) \in [-1.0, 0.6]$. The logarithmic
      profile is computed using the fourth-order Savitzky–Golay
      smoothing algorithm over the 15 nearest bins
      \citep{savitzky64}. The thick orange line indicates the slope of the stacked
      median profile, and the coloured lines indicate the density
      profiles along different intervals in position angle. Ten intervals
      are equally spaced in $\mathrm{cos}(\theta)$. For pairs of haloes, the position
      angle is defined relative to the vector joining the two haloes ($\mathrm{cos}(\theta) = \underline{r} \boldsymbol{\cdot} \underline{r}_{\rm pair}$, so $\mathrm{cos}(\theta) =1$ is directly towards the
      neighbouring halo). For the isolated haloes, the
      position angle is arbitrary (we take
      $\mathrm{cos}(\theta) = x/r$). The dotted vertical lines show the minimum,
      defined as $R_{\rm Caustic}$, of the logarithmic slope profile
      in each position angle interval. The adjacent panels show $R_{\rm Caustic}$ as a
      function of position angle. Here, the colours of the filled
      square symbols correspond to the coloured lines. For isolated
      haloes, the minima are fairly constant; however,
      $R_{\rm Caustic}$ can vary significantly for paired haloes
      (between $0.6r_{\rm 200m}$ towards/away from the companion, and
      $1.4r_{\rm 200m}$ perpendicular to the companion). This shows
      that the presence of a companion affects the outer caustic
      (often called the ``splashback'' radius) of dark matter haloes.}
          \label{fig:dens_cosang}
\end{figure*}

We identify the ``edges'' of Milky Way-mass haloes in the ELVIS,
APOSTLE and Auriga simulations using both the density and the radial
velocity profile as a function of radius. The former is motivated by
the work by \cite{diemer14}, who used the slope of the logarithmic
density profile to identify the outer edges of dark matter
haloes. Here, we apply a similar formalism, but also apply this to the
stars and subhaloes. We use the radial velocity profiles in a similar
manner.

Throughout this work we give radii in units of $r_{\rm 200m}$, defined
as the radius at which the density of a halo falls to 200 times the
universal \textit{matter} density at $z=0$
($\rho_m = \Omega_m \rho_{\rm crit}$). We also give radial velocities in units of $v_{\rm 200m}$, where $v_{\rm 200m}=\sqrt{GM_{\rm 200m}/r_{\rm 200m}}$. \cite{diemer14} show that
$r_{\rm 200m}$, rather than the commonly used $r_{\rm 200c}$, is a
more natural choice to scale haloes at large radii. However, as we
will show, $r_{\rm 200c}$ (or even $r_{\rm vir}$, \citealt{eke98,
  bryan98}) may be a more appropriate choice to define the edges of Milky
Way mass haloes. Note, for a typical NFW profile with concentration, $c=10$, $r_{\rm 200m} \approx 1.6 r_{\rm 200c}$.

\subsection{Dark Matter}
\label{sec:dm}

We first focus on the dark matter profiles of the haloes. For the
radial density profiles we use 40 evenly spaced bins in
$\mathrm{log}\left(r/r_{\rm 200m}\right)$ between $-1.0$ and
$0.6$. The logarithmic slope profile,
$\mathrm{d\,log}(\rho)/\mathrm{d\,log}(r)$, is computed using the
fourth-order Savitzky–Golay smoothing algorithm over the 15 nearest
bins \citep{savitzky64}. This choice of smoothing length allows us to
identify the strongest features in the profile, and removes most of
the noise (cf. \citealt{diemer14}). The significance of the logarithmic slope profile for dark matter haloes is discussed in detail in \cite{diemer14}.  For quiescent Milky Way mass haloes, the profile has a slowly steepening slope out to $\sim \! r_{\rm 200m}$, and then flattens to a slope of $-1$ at larger radii as the halo approaches the 2-halo term of the halo-mass correlation function \citep[e.g.][]{hayashi08}, where it is dominated by particles in different haloes. The transition between steepening and flattening results in a pronounced ``dip'' in the logarithmic slope profile (see below).

\begin{figure*}
  \centering
        \includegraphics[width=\textwidth,angle=0]{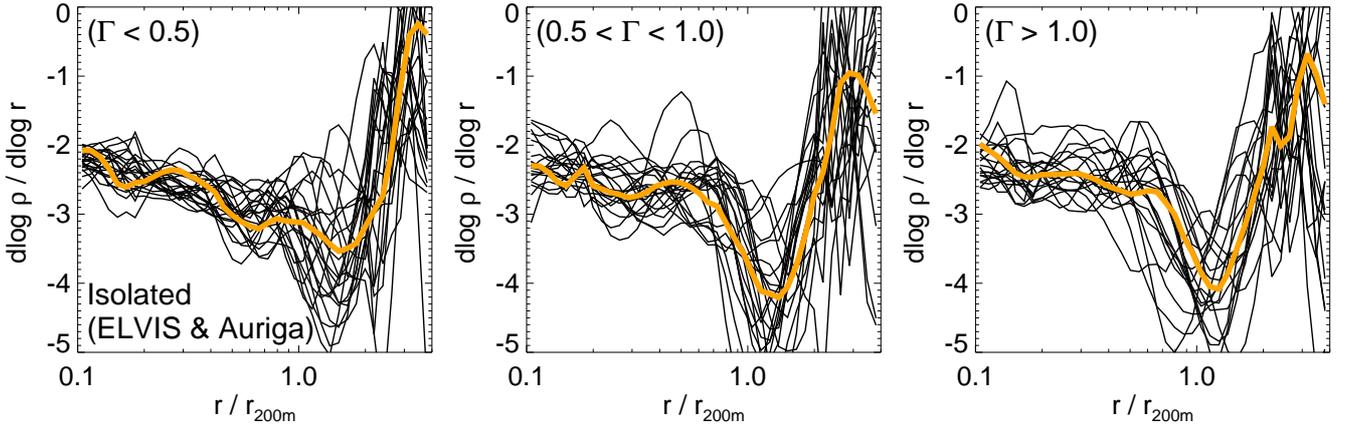}
        \caption[]{The logarithmic slope profile,
      $\mathrm{d\,log}(\rho)/\mathrm{d\,log}(r)$, of the dark matter
      density profiles for the isolated ELVIS and Auriga haloes. Here, we show three bins of recent mass accretion rate, $\Gamma$, increasing from left to right. The black lines show individual halo profiles, and the thick orange line indicates the logarithmic slope of the median density profile for each mass accretion rate bin. The feature we have termed the second caustic, which is a less pronounced than the splashback radius and is located at smaller radii, becomes more evident for low mass accretion rates.}
          \label{fig:iso_gamma}
\end{figure*}

First, we consider stacked density profiles of each simulation suite (ELVIS,
APOSTLE, and Auriga) at various position angles.  We split each halo
into intervals in position angle (0.2 width in
$\mathrm{cos}(\theta)$, see below) and compute the radial density profile in each interval. We then calculate the median stacked density profile in each interval for
the entire halo sample.  For the paired haloes, the position angle
is defined relative to the vector joining the two haloes, $\mathrm{cos}(\theta) = \underline{r} \boldsymbol{\cdot} \underline{r}_{\rm pair}$. Thus, 
$\mathrm{cos}(\theta) =1$ is directly towards the neighbouring
halo. For isolated haloes, this position angle is arbitrary and
we define $\mathrm{cos}(\theta) = x/r$, where the axes $x$, $y$ and $z$ are randomly chosen in the simulation box. In Fig.~\ref{fig:dens_cosang} we show the slopes of the median
stacked profiles. The different coloured lines show ten equally spaced intervals in $\mathrm{cos}(\theta)$, and the thick orange line shows the logarithmic slope profile of the median density profile over all position angles. For the logarithmic slope profile of the median density profile (thick orange line) we take the median density in each radial bin (over all haloes and position angles) and then compute the logarithmic slope profile. This is not the same procedure as taking the median of logarithmic slope profiles for each position angle (shown with the coloured lines), so the median profile does not always lie in the middle of these lines. The same procedure is used in subsequent plots when we show the slope profile of the median density. The dotted vertical
lines indicate the most prominent minima of
$\mathrm{d\,log}(\rho)/\mathrm{d\,log}(r)$ for each position
angle. Note these minima are chosen to have
$\mathrm{d\,log}(\rho)/\mathrm{d\,log}(r) < -2.5$ to minimize the
effect of noise. The location of these minima, $R_{\rm Caustic} / r_{\rm 200m}$, which we use to define the caustics, are shown as a function of position angle in the adjacent panels. 
Note that although we show stacked profiles over several haloes, the profiles in each position angle interval are subject to the effects of substructure. When averaging over all position angles, we can account for this (see below). However, here we explicitly check that removing substructures from the analysis does not significantly affect the results.

Previous work \citep[e.g.][]{adhikari14, diemer14, more15, diemer17}
has used the location of these minima, $R_{\rm Caustic}$, in dark
matter haloes to define the so-called ``splashback'' radius, which is
predicted in spherical models of secondary collapse
\citep[e.g.][]{fillmore84, bertschinger85}. For isolated haloes
(ELVIS-Iso, Auriga) the location of this caustic shows little
variation with position angle and is typically located at
$1.4r_{\rm 200m}$. The location of this feature is in good agreement with the location of the outermost caustic (splashback) measured in previous studies for Milky Way mass haloes \citep{diemer14, more15}. Note
that some variation with position angle is expected as the accretion
of dark matter is not isotropic (see e.g. \citealt{mansfield17});
however, as the definition of $\mathrm{cos}(\theta)$ is arbitrary for
isolated haloes, we do not expect to see large differences in the
stacked profiles. 

The location of the minimum in the paired haloes is less clear than in
the isolated haloes. Here there is more variation in
$R_{\rm Caustic}$, and the overall median stacked profile (solid
orange line) appears to have two minima (see below). The variation in
the location of $R_{\rm Caustic}$ is not random. For position angles
directly towards and away from the neighbouring halo $R_{\rm Caustic}$
is significantly smaller ($R_{\rm Caustic} / r_{\rm 200m} \sim 0.6$)
than in other directions. It is unsurprising that the caustic
\textit{towards} the neighbour is affected: here, the typical
splashback radius ($\sim \! \! 1.4r_{\rm 200m}$) runs into the
neighbouring halo. However, it is less obvious why the directly
opposite direction should be affected. For paired haloes the dynamics
of the particles are governed by the effective potential of the two
massive haloes, and there is a ``saddle point'' in the potential at
$\mathrm{cos}(\theta) =1$. Our interpretation is that along this direction particles can only
accrete from a limited distance due to the presence of the
neighbour. This material will then have less time to accelerate before
it reaches apocentre due to its smaller starting distance, and thus
will reach a smaller apocentre on the opposite side (i.e. at
$\mathrm{cos}(\theta) =-1$). Another possibility is that distribution of mass in the $\mathrm{cos}(\theta)=-1$ direction is due to the Lagrange points of the effective potential that are expected in that direction. In this scenario, particles that go beyond the Lagrange points of the effective potential escape, and at $\mathrm{cos}(\theta)=-1$ we are seeing a feature shaped by the presence of a such a Lagrange point, which is closer than it would be for an isolated halo.

\begin{figure*}
    \begin{minipage}{0.45\linewidth}
        \centering
        \includegraphics[width=\textwidth,angle=0]{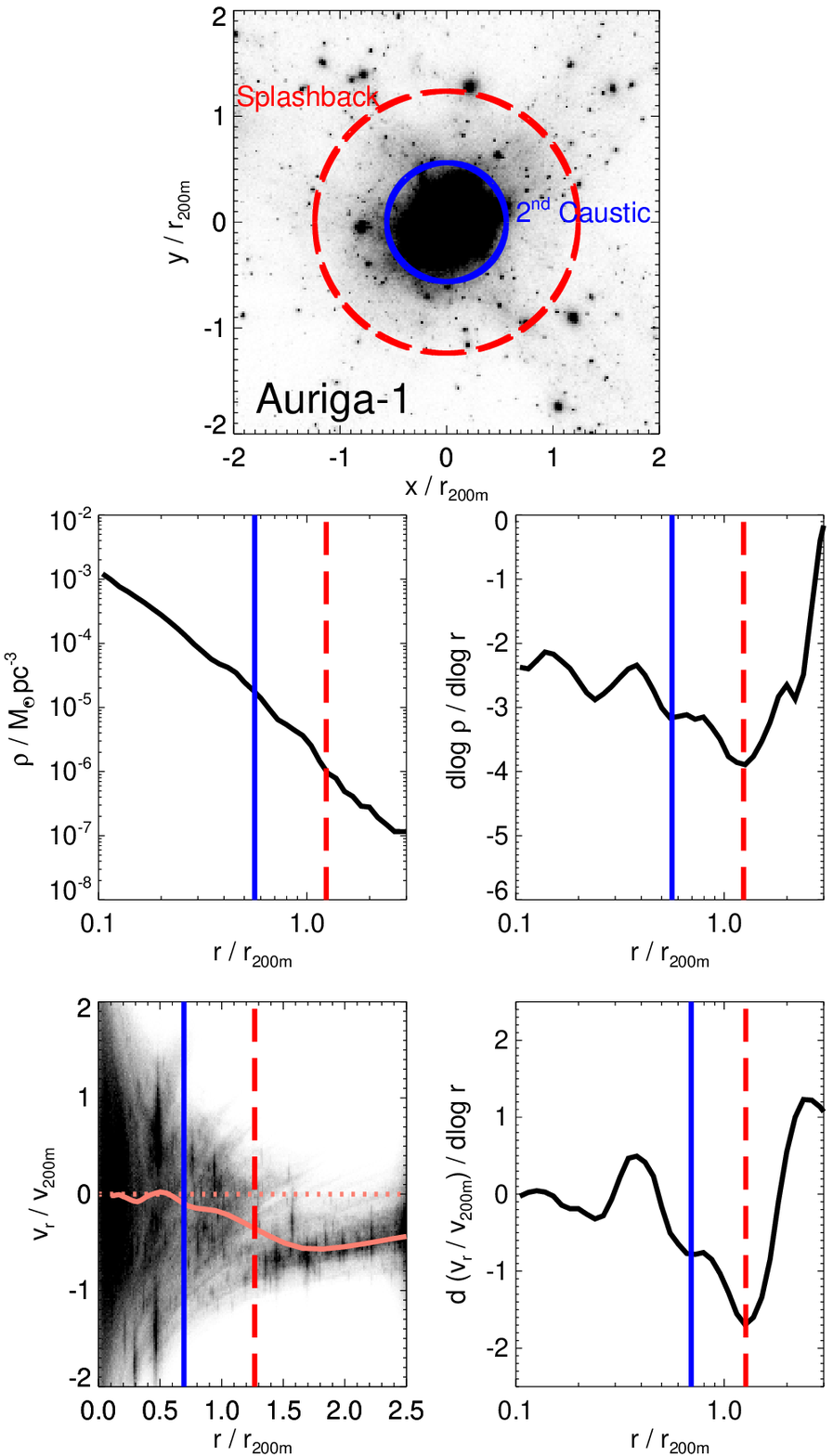}
    \end{minipage}
    \hspace{10pt}
    \begin{minipage}{0.45\linewidth}
        \centering
        \includegraphics[width=\textwidth,angle=0]{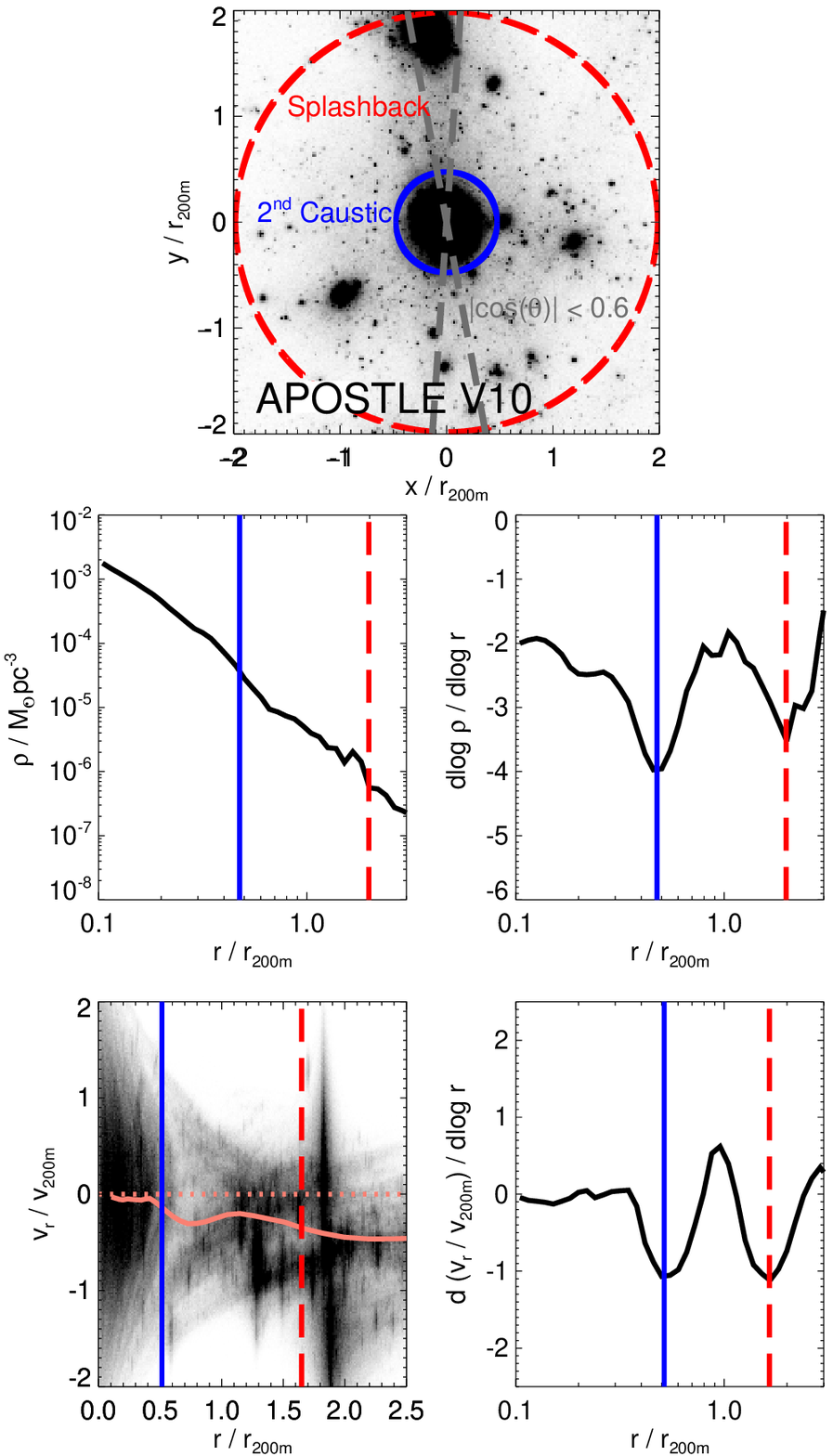}
    \end{minipage}
        \caption[]{Two example haloes from Auriga (left panels) and
          APOSTLE (right panels). Here we show the density of dark
          matter in $(x,y)$ projection (top panels), the radial
          density profiles (middle panels), and the radial velocity
          profiles (bottom panels). The shading in the top and bottom
          (left) panels shows $200\times200$ pixels saturated at the
          95th percentile of the 2D histogram. In addition to the
          density and radial velocity profiles, we also show the
          logarithmic slope profiles of these quantities:
          $\mathrm{d\,log}(\rho)/\mathrm{d\,log}(r)$ and
          $\mathrm{d}\,(v_r)/\mathrm{d\,log}(r)$. These logarithmic
          slope profiles are used to identify caustics in the dark
          matter. The vertical lines indicate the splashback radius
          (red dashed lines) and the second caustic (blue solid
          line). These radii, computed from the density profile, are also shown in the top plots. The
          position angles excluded in the paired haloes to compute these
          quantities is shown in the top-right panel
          ($|\mathrm{cos}(\theta)| < 0.6$). The radial velocity
          profiles (the solid pink lines show the median profile, and the dotted pink line indicates the zero level for reference) suggest that the splashback radius is related to the
          material infalling onto the haloes for the first time, and
          the second caustic relates to the edge of the virialized
          material, which has undergone at least two orbital passages
          through pericentre. The caustics defined in density or
          velocity space are closely related, albeit with some scatter
          (see Fig \ref{fig:dm_caustics}).} 
          \label{fig:dm_eg}
\end{figure*}

\begin{figure*}
    \begin{minipage}{\linewidth}
        \centering
        \includegraphics[width=\textwidth,angle=0]{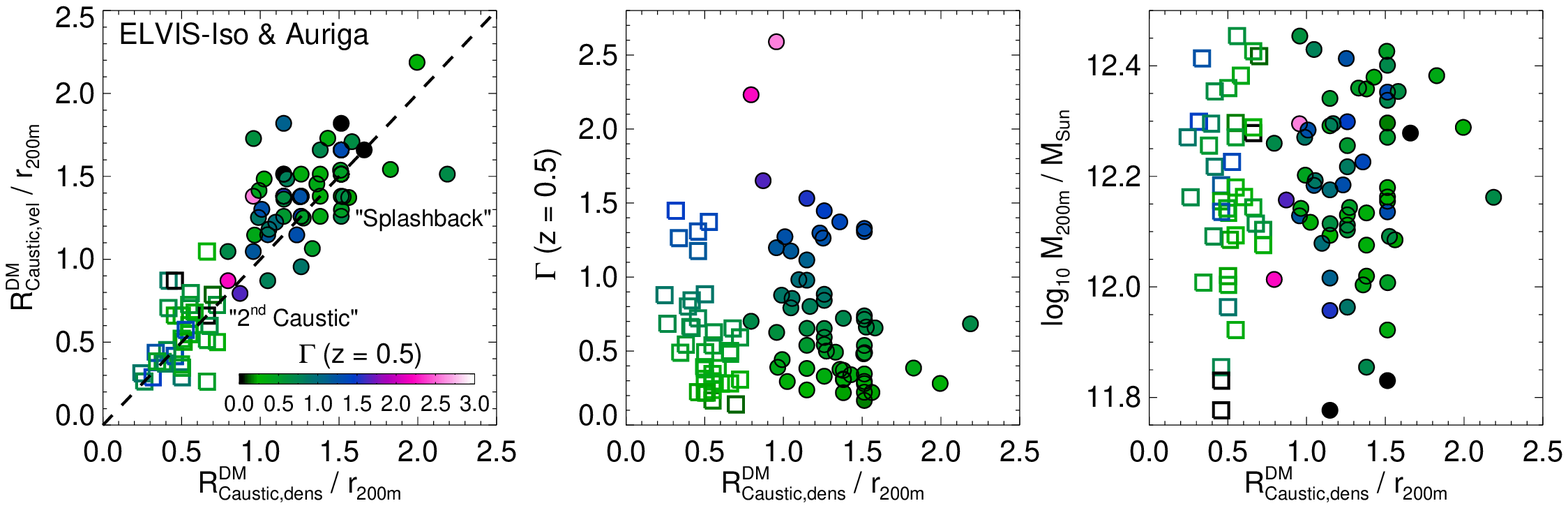}
    \end{minipage}
    \hspace{10pt}
    \begin{minipage}{\linewidth}
        \centering
        \includegraphics[width=\textwidth,angle=0]{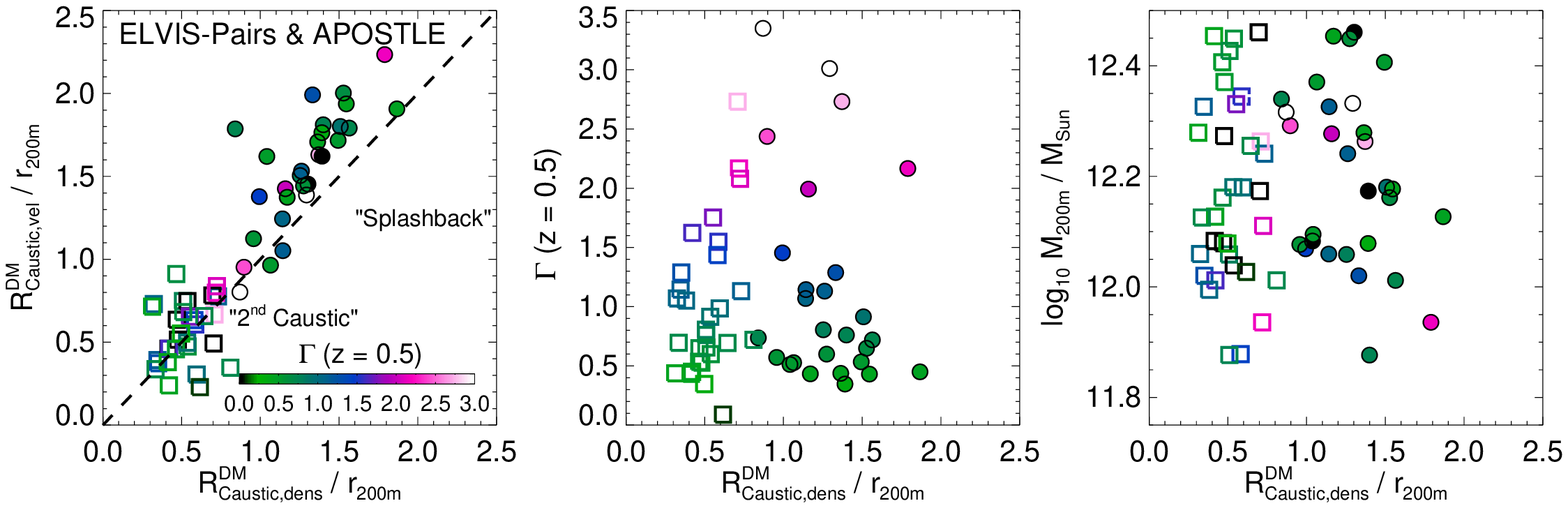}
    \end{minipage}
    \caption[]{The dark matter caustics of individual haloes in
      isolated (top panels) and paired (bottom panels)
      environments. The left-hand panels show the position of the
      density caustics against the radial velocity caustics. The
      filled circles indicate the splashback radius and the open
      squares the second caustics. The dashed lines show the
      one-to-one relation. The splashback radii are more poorly
      defined in the paired haloes (e.g. in $\sim \!  \!20$ percent of the
      paired haloes a splashback radius cannot be cleanly
      identified). However, the properties of the second caustics are
      similar between paired and isolated haloes. The symbols are
      coloured according to the mass accretion rate,
      $\Gamma(z=0.5)$. For haloes with rapid recent accretion the
      splashback radius tends to be smaller, and closer to the second
      caustic. Indeed, most cases in which \textit{two} caustics could
      not be clearly identified have relatively high $\Gamma$. The
      middle panels show the mass accretion rate against the radius of
      the dark matter caustics and the right-hand panels show the
      ($z=0$) halo masses against the dark matter caustics.}
          \label{fig:dm_caustics}
\end{figure*}

The location of a second caustic at
smaller radii has been seen in previous work (see e.g. figs 10, 13, 14
in \citealt{diemer14}) and has been demonstrated explicitly in
\cite[][see their fig. 9]{adhikari14}. \cite{adhikari14} show that for
slowly accreting haloes, the stream of splashback material is separated
from the rest of the virialized matter in the halo, and the location
of the second caustic becomes more pronounced. The majority of Milky
Way-mass haloes are slowly accreting (especially relative to
cluster-sized haloes), so it is particularly intriguing that we detect this feature here. Curiously, the typical location of this second caustic
corresponds to $r_{\rm 200c}$, rather than $r_{\rm 200m}$ (as
$r_{\rm 200m} \sim 1.6 \times r_{\rm 200c}$). We first noted this secondary feature in the paired haloes, however, this feature is also apparent in the individual profiles of the isolated haloes (see below). This feature can be difficult to see in the stacked profiles in Fig. \ref{fig:dens_cosang} as there is considerable halo-to-halo scatter, and the signal is relatively weak (especially relative to the splashback radius for isolated haloes). In Fig. \ref{fig:iso_gamma} we show the logarithmic slope profiles for individual haloes in the isolated ELVIS and Auriga runs. Here, we separate the haloes into three bins with increasing (recent) mass accretion rate from left to right. The thick orange lines show the logarithmic slope profile of the median density profiles in each bin (where the differential profile is computed after finding the median density in each radial bin, as described above). We use the definition given by \cite{diemer14} to define mass accretion rate:
\begin{equation}
\label{eq:gamma}
  \Gamma = \frac{\mathrm{log}M_{\rm vir}(z_1)-\mathrm{log}M_{\rm vir}(z_2)}{\mathrm{log}(a_1)-\mathrm{log}(a_2)}
\end{equation}
where $z_1=0$ and $z_2=0.5$.  Note
when computing the individual halo profiles we compute the
median value over 10 equally spaced intervals in position angle (i.e. 0.2 width in $\mathrm{cos}(\theta)$) for each radial bin. This procedure has the advantage of minimizing the effect of substructure in the profile \citep{mansfield17}. We have checked that explicitly removing (bound) substructures produces very similar results, however we do caution that there are other inhomogeneities present in the density that could effect the results, but we expect that our procedure will account for the most prominent irregularities. Fig. \ref{fig:iso_gamma} illustrates two important points. First, as mentioned above, there is wide range in halo-to-halo scatter, particularly, for any second caustic features. Second, the second caustic becomes more prominent at lower mass accretion rates, as predicted by \cite{adhikari14}. Note that the stacked profiles, particularly at low accretion rates, hint at three separate caustics in the logarithmic density profile. The very inner ``dips'' likely correspond to the apocentres of early, massive mergers in the halo's assembly history. However, we caution against over-interpretation of these features as they can have low significance. Finally, it is worth noting that, although we see evidence for a second caustic in both paired and isolated haloes, it is not necessary true that the origin of the caustic is the same in both cases. Indeed, there could be multiple, interconnected causes for this interesting feature in Galactic-sized haloes. We now explore the second caustic feature further by analysing individual haloes in more detail.

In Fig. \ref{fig:dm_eg} we show two example haloes. The left panels
show the dark matter distribution of Auriga-1 (an isolated halo), and
the right panels show APOSTLE V10 (a paired halo: in
Fig. \ref{fig:dm_eg} the coordinate system is centred at
$(x,y,z)=(61.948, 24.230, 48.305)$ Mpc in the V10 system, see Table A1
in \citealt{fattahi16}). The top panels show a 2D projection of the
dark matter distribution, the middle panels the density profile
and logarithmic slope profile, and the bottom panels the radial
velocity profile and corresponding logarithmic slope profile.  The dashed red lines indicate the splashback radius and the solid blue line the second caustic. For the paired haloes, caustics are
identified by excluding position angles with
$|\mathrm{cos}(\theta)| > 0.6$. The second caustic is
located at a smaller radius and is less pronounced than the splashback
radius. We generally find that the second caustics are easier to
identify in the individual halo density profiles, than in the stacked
profiles (see e.g. Figs \ref{fig:dens_cosang} and \ref{fig:iso_gamma}). This is likely because
the feature is relatively weak and gets smeared out over a range of
radii when the profiles are stacked together. 

The second caustic can also be seen in the radial velocity
profile. Here, we use the local minimum of
$\mathrm{d}(v_r)/\mathrm{d\,log}(r)$ to identify the caustics. The
velocity and density caustics typically align on average, but there is some
scatter (see Fig. \ref{fig:dm_caustics}). The radial velocity profile
allows us to see more clearly what the second caustic is. The feature looks similar to the second caustic features shown in \cite{adhikari14}, and we suggest that this feature relates to the edge of
the material in the halo at the position where particles have
completed at least two passages through pericentre. The splashback
radius is located where material is outgoing for the first time, and
particles have only completed one pericentric passage. The existence
of two caustics, each defining different regions of the halo, begs the
question: which should we use to define the edge of the halo? This
question is particularly relevant for low mass accreting haloes, where
the splashback and second caustic are well separated
\citep[][]{adhikari14}. Our Milky Way is located in the Local Group
and neighbours a massive halo, so the definition of splashback radius
is less clear (and indeed overlaps with the halo of M31). For this
reason, we suggest that the most meaningful radius for the Milky Way
is the second caustic. We will show in Section~\ref{sec:stars} that
this definition is also applicable to the stellar material. Note, however, that although we have defined this interesting feature as the ``second caustic'', this does not necessarily correspond to the classical definition of second caustic from spherical (or ellipsoidal) collapse models (as seen in \citealt{adhikari14}). In particular, the wide halo-to-halo scatter, and the apparent correlation with the stellar distribution (see following section), could point to a merger origin, i.e. from the apocentre of the last major merger. In addition, we caution that the second caustic, as we have defined it in this work, could have multiple origins that vary from halo-to-halo. The actual origin of this feature will require further investigation, ideally with particle evolution tracking.

\begin{figure*}
    \begin{minipage}{0.49\linewidth}
        \centering
        \includegraphics[width=\textwidth,angle=0]{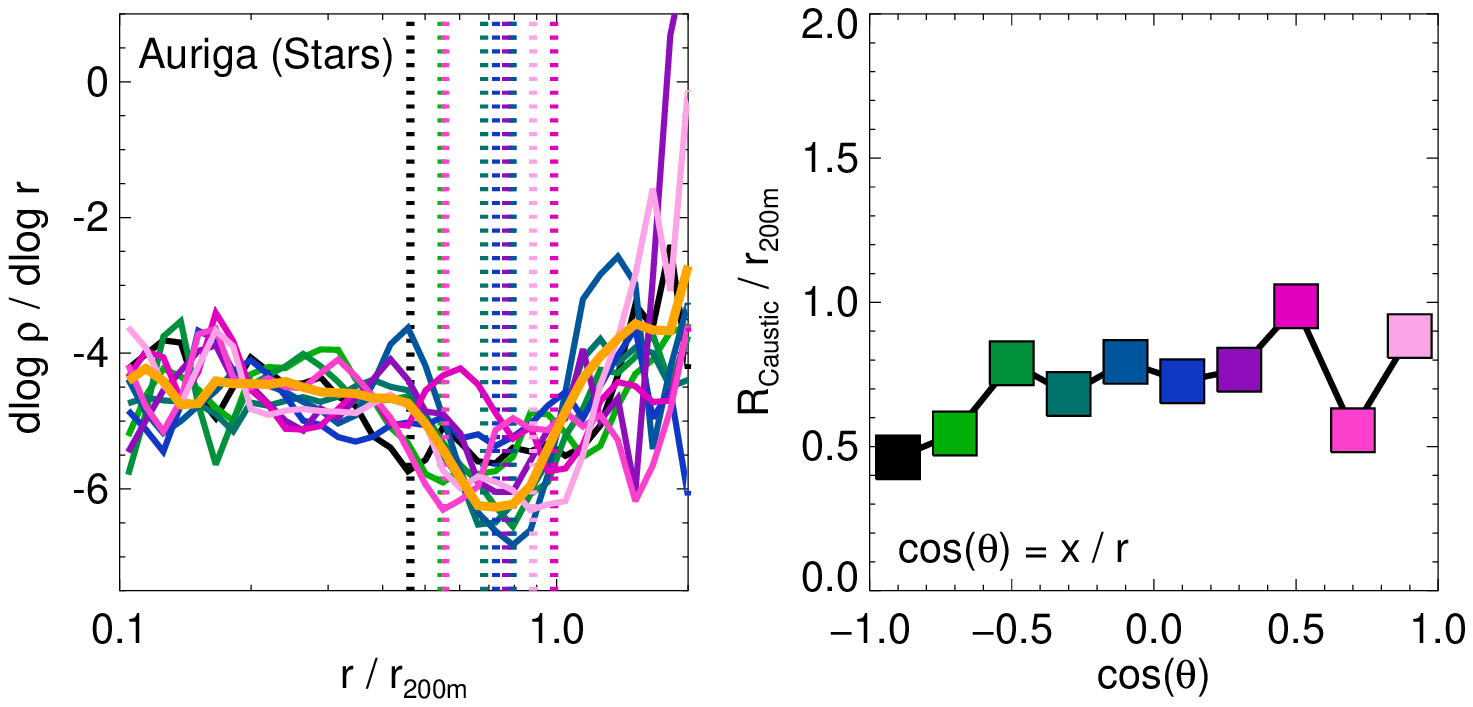}
    \end{minipage}
    \begin{minipage}{0.49\linewidth}
        \centering
        \includegraphics[width=\textwidth,angle=0]{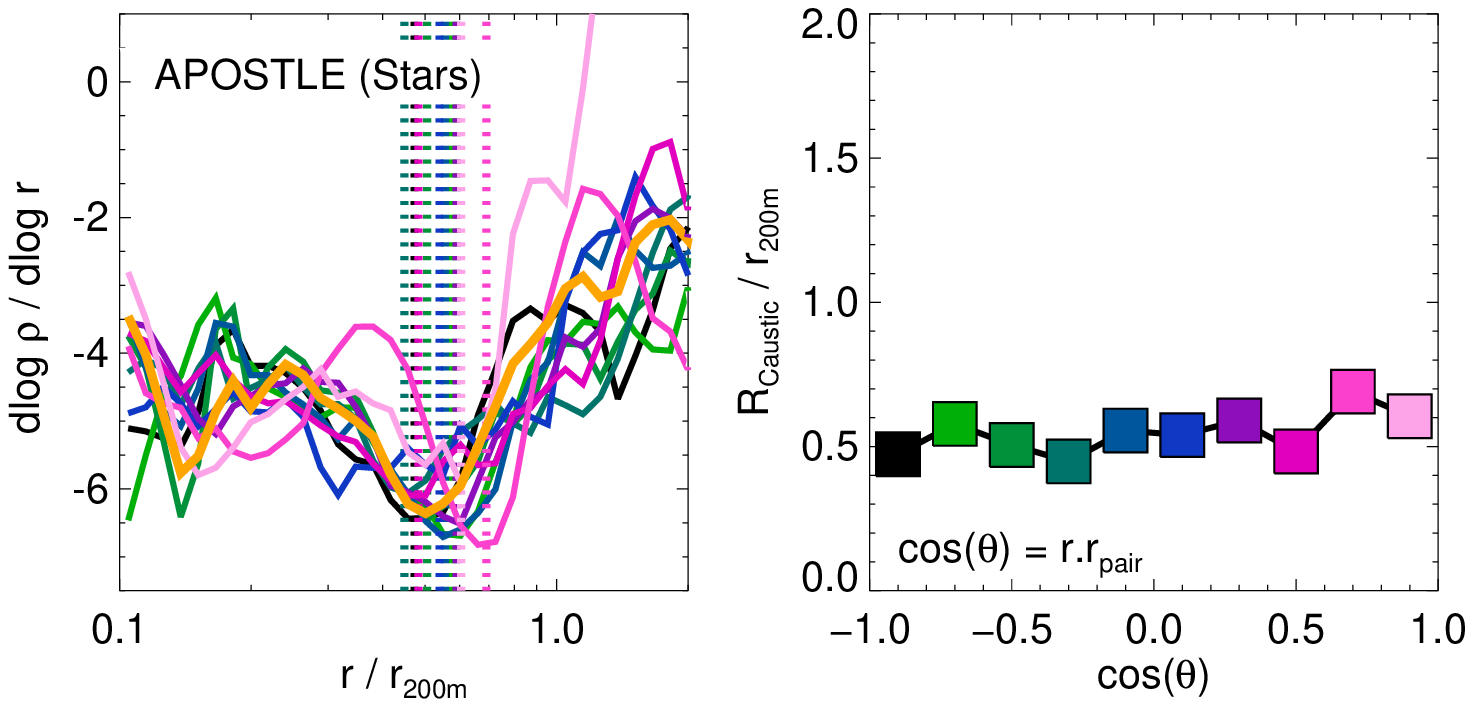}
    \end{minipage}
    \caption[]{The logarithmic slope profile,
      $\mathrm{d\,log}(\rho)/\mathrm{d\,log}(r)$,  of the stellar
      density profiles of the Auriga (left) and APOSTLE (right)
      haloes. Here, 40 evenly space bins in $\mathrm{log}(r/r_{\rm
        200m})$ have been used in the range $\mathrm{log}(r/r_{\rm
        200m}) = [-1.0, 0.6]$. The logarithmic profile is computed
      using the fourth-order Savitzky–Golay smoothing algorithm over
      the 15 nearest bins \citep{savitzky64}. The thick orange line
      indicates the logarithmic slope profile of the median density profile, and the coloured lines
      show the slope profiles along different intervals in position
      angle. Ten intervals
      are equally spaced in $\mathrm{cos}(\theta)$. For pairs of haloes, the position angle is defined
      relative to the vector joining the two haloes ($\mathrm{cos}(\theta) = \underline{r} \boldsymbol{\cdot} \underline{r}_{\rm pair}$, so $\mathrm{cos}(\theta) =1$ is directly towards the
      neighbouring halo). For the isolated haloes, the position
      angle is arbitrary (we take $\mathrm{cos}(\theta) =
      x/r$). The dotted vertical lines show the minimum,
      defined as $R_{\rm Caustic}$, of  the logarithmic slope profile
      in each position angle interval. The adjacent panels show $R_{\rm Caustic}$ as a
      function of position angle. The colours of the filled
      squares correspond to the coloured lines. The caustics
      for paired and isolated haloes are similar, and are typically
      located at $0.6r_{\rm 200m}$.} 
          \label{fig:stars_cosang}
\end{figure*}

In Fig. \ref{fig:dm_caustics} we show the positions of the dark matter
caustics for individual haloes in isolated (top panels) and paired
(bottom panels) environments. The caustics are identified as minima in
the $\mathrm{d\,log}(\rho)/\mathrm{d\,log}(r)$ and
$\mathrm{d}(v_r)/\mathrm{d\,log}(r)$ profiles. We consider the two
most prominent (outer) caustics, and only consider features with
$\mathrm{d\,log}(\rho)/\mathrm{d\,log}(r) < -2.5$ and
$\mathrm{d}(v_r)/\mathrm{d\,log}(r) < -0.25$, respectively. In
addition, for every individual halo we visually inspect the profiles
to ensure we are not confusing noise with a real caustic. The left
panels show the position of the velocity caustics against the density
caustics.  The filled circles show the splashback radii and the open
squares the second caustics. Note that for isolated haloes the
splashback radius can be identified in almost all of the haloes; 
however, even with a restriction on position angle, this can be harder
to detect in the paired haloes. Over all paired haloes (in ELVIS and
APOSTLE) 21 percent have no detectable splashback radius in density or
velocity. Moreover, the density and velocity caustics are not as
closely aligned in the paired environments. On the other hand, the
detection efficiency of the second caustic is very similar between
isolated and paired haloes of similar mass (e.g. by comparing ELVIS Isolated and Paired haloes). There is no discernible second caustic in
16 percent of the haloes (over all haloes in ELVIS, APOSTLE and Auriga), and the non detections are typically more
massive haloes with higher recent accretion rates (see below and Fig. \ref{fig:iso_gamma}). The detected second caustics range in radii between $0.3-0.8r_{\rm 200m}$ and have density slopes at these radii of $\sim -2.5$ to $-4.5$.

The symbols in Fig.~\ref{fig:dm_caustics} are coloured according to
the recent mass accretion rate (see Eqn. \ref{eq:gamma}).  The majority of haloes have quite low
recent mass accretion rates ($\Gamma < 1$), as expected for Milky Way
mass haloes. The middle panels of Fig.~\ref{fig:dm_caustics} show how
the positions of the caustics relate to $\Gamma$.  The caustics in the
isolated haloes are typically at smaller radii for haloes with higher
recent mass accretion rates (as shown in \citealt{diemer17} over a
wider mass range). However, this trend is not present in the paired
environments, particularly for $\Gamma > 1.5$. This is likely because
the splashback radius and the second caustic run into each other at
higher mass accretion rates, and are harder to
distinguish. Furthermore, $\Gamma$ is poorly defined in paired
environments where the outer profiles of the neighbouring haloes
overlap. Finally, we show the location of the caustics as a function
of halo mass in the right-hand panels. We see very little dependence between $R_{\rm Causitc}/r_{\rm 200m}$ and halo mass. Indeed, analytical models predict that mass accretion
rate, rather than halo mass, is the more important physical quantity
the determines the splashback radius \citep[e.g.][]{adhikari14}.

\subsection{Stars}
\label{sec:stars}

\begin{figure}
    \begin{minipage}{0.45\linewidth}
        \centering
        \includegraphics[width=\textwidth,angle=0]{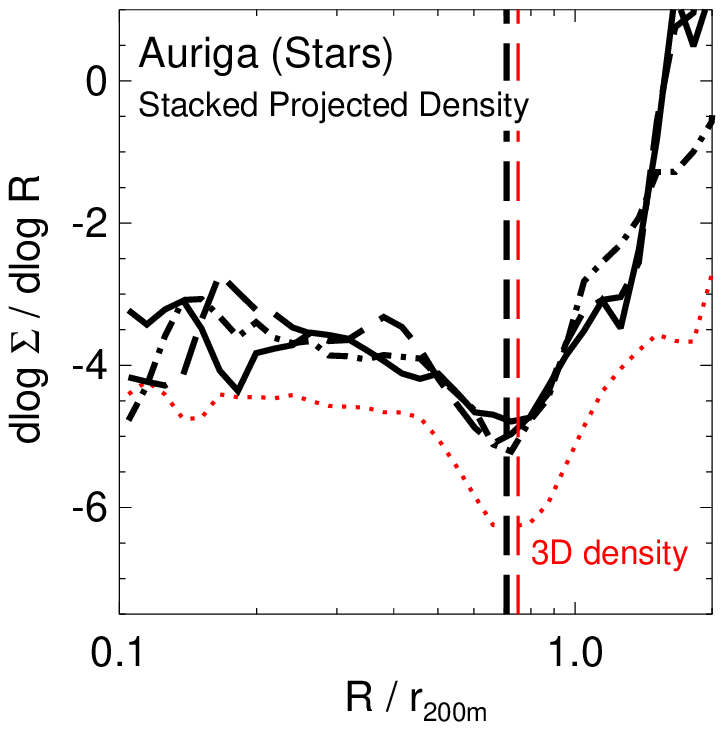}
    \end{minipage}
    \hspace{10pt}
    \begin{minipage}{0.45\linewidth}
        \centering
        \includegraphics[width=\textwidth,angle=0]{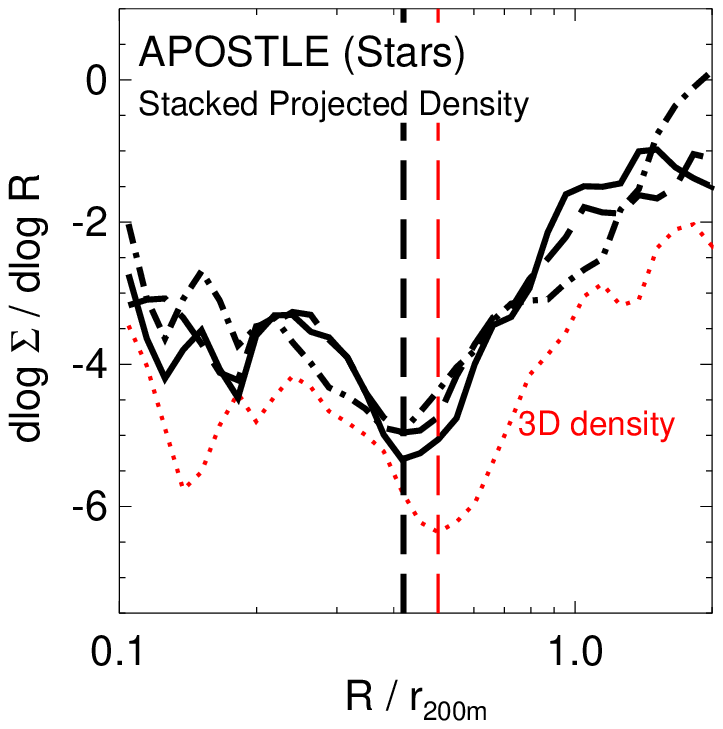}
    \end{minipage}
    \caption[]{The logarithmic slope profile,
      $\mathrm{d\,log}(\Sigma)/\mathrm{d\,log}(R)$, of the stellar
      surface density profiles of  the Auriga (left) and APOSTLE
      (right) haloes. Here, 40 evenly spaced bins in
      $\mathrm{log}(R/r_{\rm 200m})$ have been used in the range
      $\mathrm{log}(R/r_{\rm 200m}) = [-1.0, 0.6]$. The logarithmic
      profile is computed using the fourth-order Savitzky–Golay
      smoothing algorithm over the 15 nearest bins
      \citep{savitzky64}. The three linestyles show the stacked
      profiles for three (random) projections. For comparison, the
      logarithmic slope profile of the 3D stellar density is shown
      with the dotted red line (see Fig.~\ref{fig:stars_cosang}). A
      well-defined edge is also seen in the (stacked) projected
      stellar density profiles, although this is a weaker feature than
      in the 3D case.}
        \label{fig:star_proj}
\end{figure}

\begin{figure*}
    \begin{minipage}{0.45\linewidth}
        \centering
        \includegraphics[width=\textwidth,angle=0]{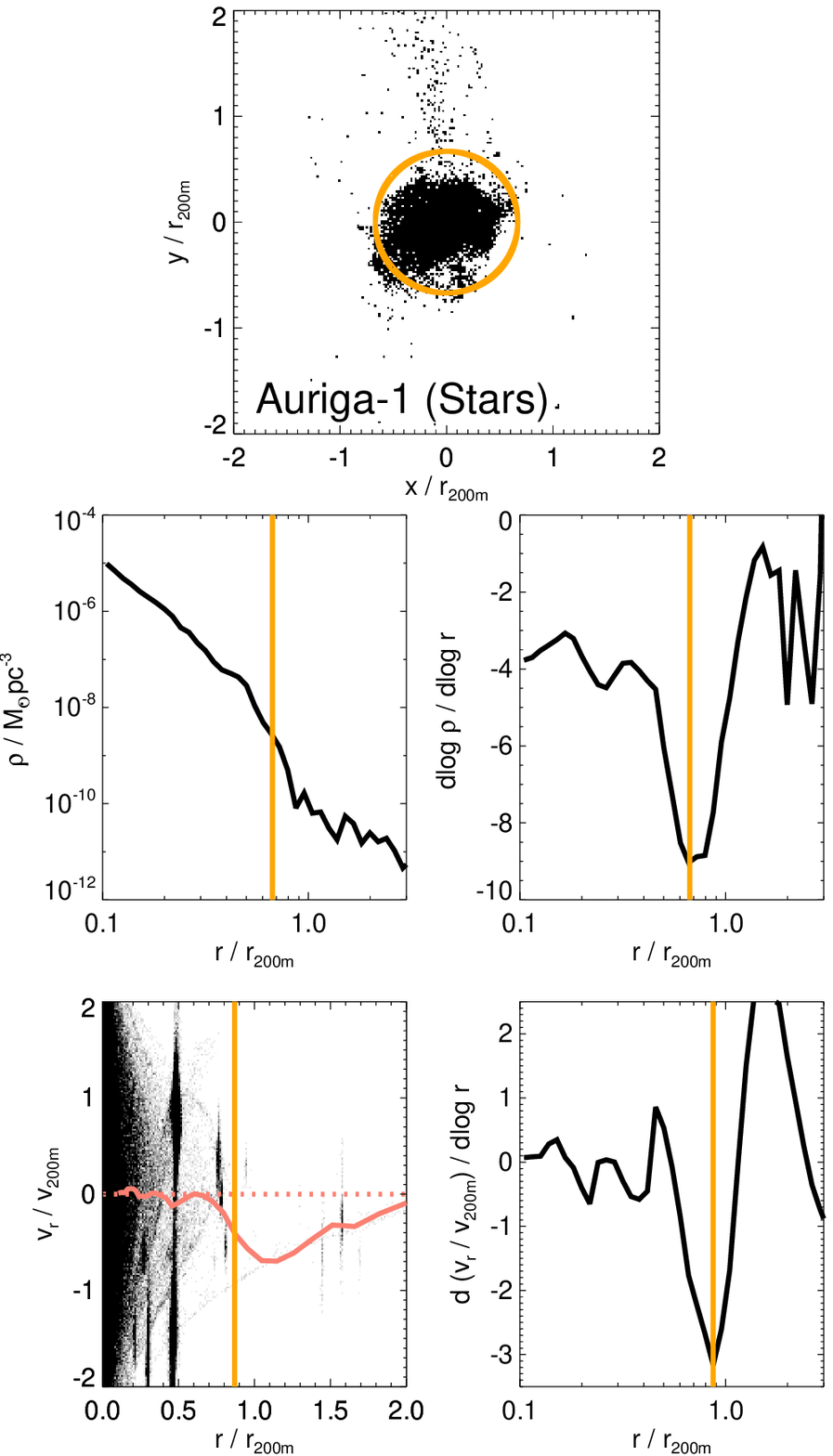}
    \end{minipage}
    \hspace{10pt}
    \begin{minipage}{0.45\linewidth}
        \centering
        \includegraphics[width=\textwidth,angle=0]{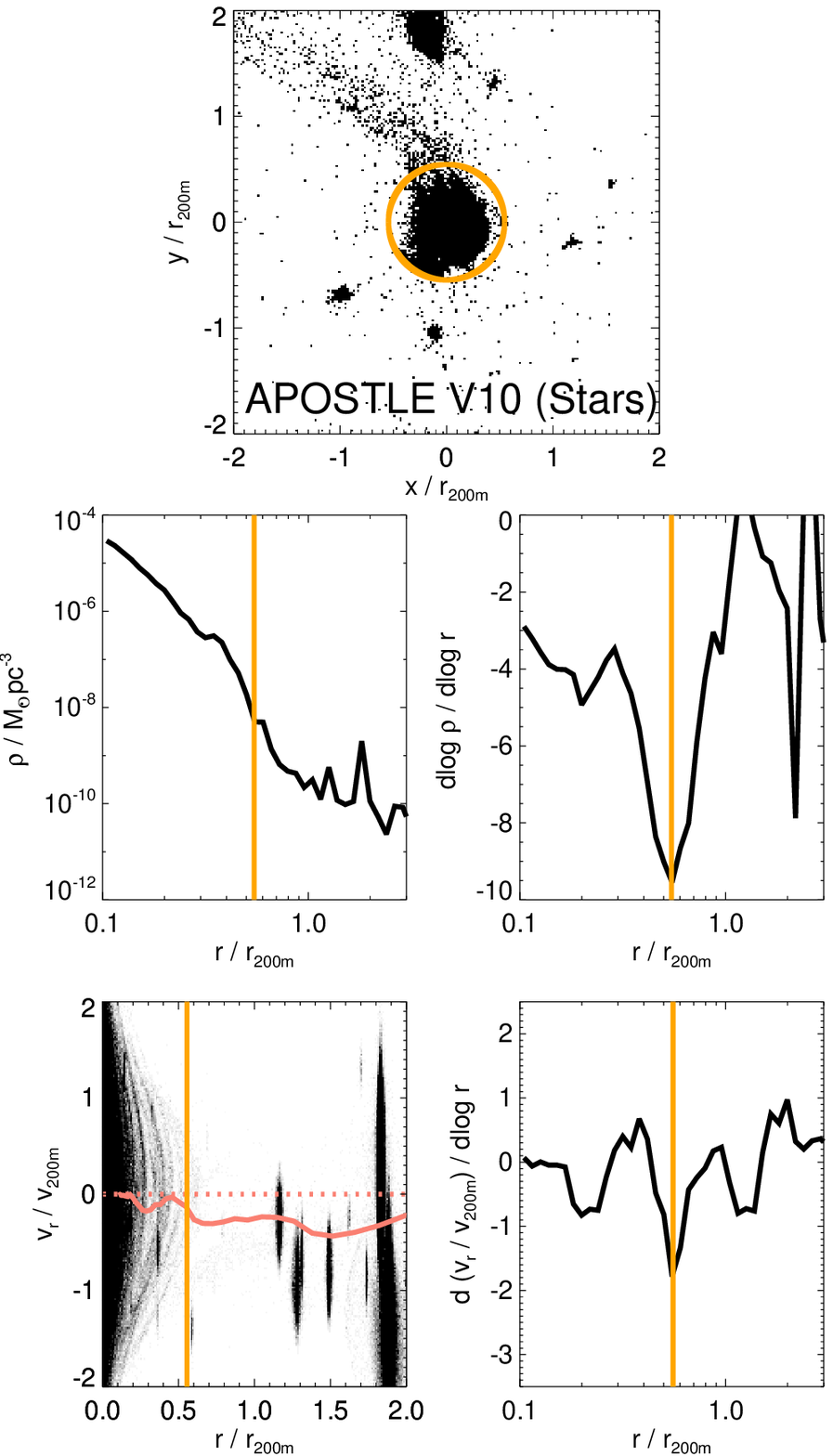}
    \end{minipage}
    \caption[]{Two example haloes from Auriga (left panels) and
      APOSTLE (right panels). These are the same haloes shown in
      Fig.~\ref{fig:dm_eg}. Here, we show the density of stars in the
      $(x,y)$ projection (top panels), the radial density profiles
      (middle panels), and the radial velocity profiles (bottom
      panels). The shading in the top and bottom (left) panels shows
      $200\times200$ pixels saturated at the 90th percentile of the 2D
      histogram. In addition to the density and radial velocity
      profiles, we also show the logarithmic slope profiles of these
      quantities: $\mathrm{d\,log}(\rho)/\mathrm{d\,log}(r)$ and
      $\mathrm{d}\,(v_r)/\mathrm{d\,log}(r)$.  The stellar caustics
      are identified as minima in the logarithmic slope profiles, and
      are indicated with the vertical solid lines. }
        \label{fig:star_eg}
\end{figure*}

\begin{figure*}
  \centering
        \includegraphics[width=\textwidth,angle=0]{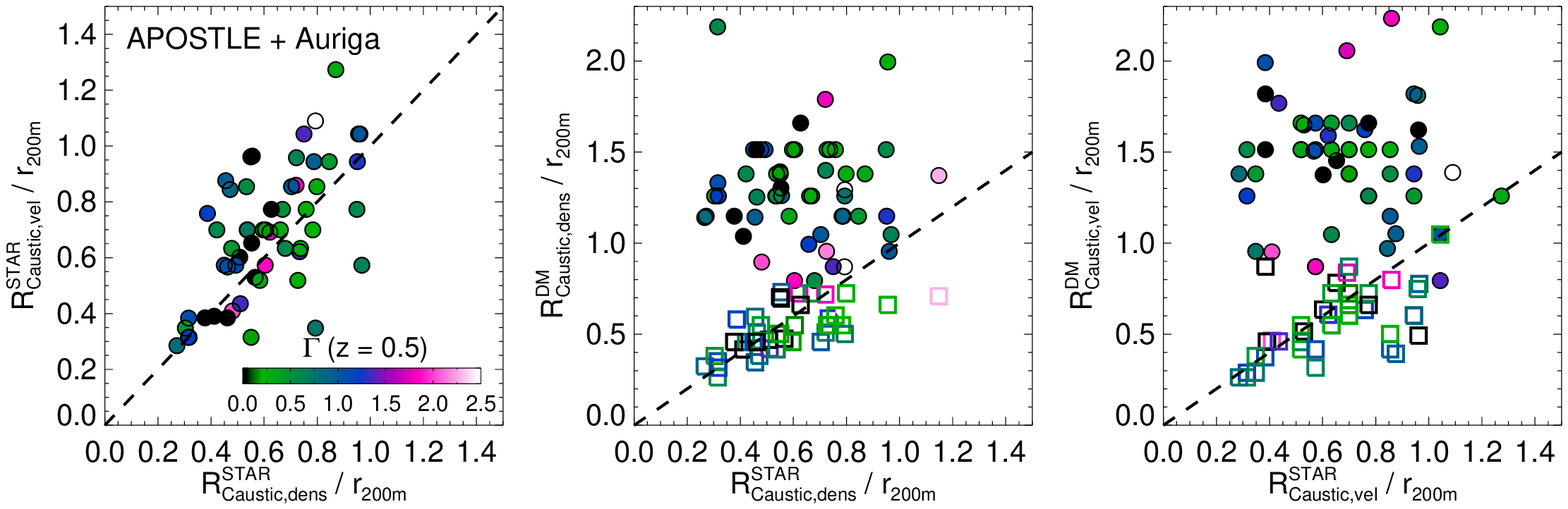}
        \caption[]{The positions of the stellar caustics in Auriga and
          APOSTLE haloes. The left-hand panel shows the radii of the
          stellar density caustics against the radii of the stellar
          radial velocity caustics. The middle and right-hand panels
          show the radii of the stellar density (middle) and velocity
          (right) caustics against those of the dark matter
          caustics. The filled circles and open squares indicate the
          dark matter splashback and second caustic radii. The symbols
          are coloured according to the mass accretion rate,
          $\Gamma(z=0.5)$. The dashed lines show the one-to-one
          relation. Note that the DM caustics at large radii appear discretized owing to the logarithmic binning. Over a wide range in radii (out to
          $\sim \! \! 0.8r_{\rm 200m}$) the stellar caustics
          correspond to the second caustic in the dark matter. In a
          few cases where $R^{\rm STAR}_{\rm Caustic}$ is large, the
          stellar caustic can lie in between the dark matter caustics,
          and can even be closer to the splashback radius. }
          \label{fig:star_caustics}
\end{figure*}

We now turn our attention to the stellar material in Milky Way-sized
haloes. We analyse the APOSTLE and Auriga simulations which
include baryonic material. In Fig. \ref{fig:stars_cosang} we show the
logarithmic slope of the stellar density profiles of the Auriga
(left) and APOSTLE (right) haloes. We use the same bin sizes and
smoothing technique as for the dark matter. As in
Fig.~\ref{fig:dens_cosang}, the median stacked profiles are shown, and
the different colours show ten different intervals in position angle. The
solid orange line shows the logarithmic slope of the median density profile for all haloes over all
position angles. We indicate the minimum in the logarithmic slope for 
each position angle with dotted lines and we only consider minima
with $\mathrm{d\,log}(\rho)/\mathrm{d\,log}(r) < -5$ (although
typically the stellar caustics are much stronger than this, and range from  $-15 < \mathrm{d\,log}(\rho)/\mathrm{d\,log}(r) < -5$).  The
location of these minima are shown as a function of position angle in
the adjacent plots (and colour coded accordingly). There is some
variation of $R_{\rm Caustic}$ with position angle, but
there is no obvious trend. It is notable that the profiles of the paired
haloes (APOSTLE) and isolated haloes (Auriga) are similar, and the
caustics are typically found at $0.6r_{\rm 200m}$. Interestingly, this
is exactly the radius that we identified in Fig.~\ref{fig:dens_cosang}
as the second caustic in the dark matter. Below, we focus on the
profiles of individual haloes, and explicitly examine this apparent
connection between the stars and dark matter.

We also show the logarithmic slope of the \textit{projected} stellar
density profiles in Fig.~\ref{fig:star_proj}. Here we show the stacked
profiles of all Auriga (left) and APOSTLE (right) haloes. This 2D
measure is relevant for stellar halo density profiles of external
Milky Way-mass galaxies for which only two spatial coordinates are
known. The three different linestyles indicate three (random)
projections, and the dotted red line shows the stacked 3D profile
(repeated from Fig.~\ref{fig:stars_cosang} for comparison). A
well-defined ``edge'' is also seen in the projected profiles. This
occurs at slightly lower radii (in projection) relative to the 3D
radius (by $\sim \! \! 0.1 r_{200m}$), and is a weaker feature than in
the 3D profiles. However, the clear detection in 2D is encouraging for
studies of external stellar haloes. Currently, surveys like Ghosts
\citep{harmsen17} and Dragonfly \citep{merritt16} are only able to
probe the stellar halo density out to $\sim \! \! 50-80$ kpc. However,
with deeper observations and future wide-field facilities such as the
Nancy Grace Roman Space Telescope \citep[][]{wfirst}, the
radial range of interest, beyond 150 kpc should be accessible for
nearby galaxies. Furthermore, the signal of the stellar edge could be
enhanced by stacking the profiles of several galaxies.

In Fig.~\ref{fig:star_eg} we show the stellar distribution of two
example haloes, Auriga-1 on the left and APOSTLE V10 on the right. The
top panels show a 2D projection, the middle panels the stellar density
and logarithmic slope profiles, and the bottom panels the radial
velocity and associated logarithmic slope profiles. The solid orange
line indicates the caustic in the stellar distribution. We typically
identify only one clear outer caustic (cf. the dark matter where we
commonly find two) at $\sim \! \! 0.6r_{\rm 200m}$.  However, there
can be less prominent caustics at smaller radii, which are associated
with apocentres of past accretion events (these can be seen in both dark matter and stars, see e.g. figs \ref{fig:dm_eg} and \ref{fig:star_eg}).  Such a feature has already
been seen in the Milky Way halo at $r \sim 20$ kpc, and is likely
related to the apocentre of the \textit{Gaia}-Sausage/Enceladus event
\citep{deason13,deason18}. In this work we are interested in the
caustic that defines the edge of the stellar material, and is hence
associated with the furthest apocentre of stars bound to the Galaxy. The radial
velocity profiles suggests that the location of this stellar caustic
coincides with the edge of the material that has completed at least
two pericentric passages, similarly to the second caustic in the dark
matter (see below).  We find no obvious difference between the
isolated and paired haloes, which is unsurprising as the location of
the stellar caustic ($0.6r_{\rm 200m}\sim r_{\rm 200c} \sim 220$ kpc)
does not generally overlap with the neighbouring halo.

In Fig.~\ref{fig:star_caustics} we examine the stellar caustics of
individual haloes in more detail. We are able to identify a stellar
caustic in over 90 percent of the haloes. Those cases where we cannot
clearly identify a feature (in either density or velocity) are
typically cases where there is very recent accretion and the outer
density profiles are messy. Note we typically only consider stellar
caustics with $\mathrm{d\,log}(\rho)/\mathrm{d\,log}(r) < -5$ or
$\mathrm{d}\,(v_r)/\mathrm{d\,log}(r) < -1.0$, which we choose to be
distinct from the noise level. The left panel of
Fig.~\ref{fig:star_caustics} relates the positions of the velocity and
density caustics of the stars. These caustics generally coincide but
there is significant scatter. The points are colour coded according to
the recent (total) mass accretion rate, $\Gamma(z=0.5)$ (see
Eqn. \ref{eq:gamma}). In the middle (density) and right-hand
(velocity) panels we relate the stellar caustics to the dark matter
caustics. Solid filled symbols are used for the splashback radii of
the dark matter and open squares for the second caustic of the dark
matter. As mentioned earlier, the stellar caustics are strongly
related to the second caustic in the dark matter. Note the dashed line
indicates the one-to-one relation; this is not a fit! This relation
holds for $\sim (0.3-0.8)r_{\rm 200m}$, but seems to break down at
larger radii. This discrepancy at large radii is likely for two
reasons. Firstly, when $R^{\rm STAR}_{\rm Caustic}$ is large the
stellar caustic can be closer to the splashback radii, or even
somewhere between the second caustic and the splashback
radius. Secondly, the stellar caustic is harder to define at very
large distances ($0.8r_{\rm 200m} \sim 300$ kpc) where the density of
stars is very low.

We leave a more thorough analysis of how the dynamics of the star
particles relate to the dark matter to future work. However, it is
worth discussing the possible origin of the correlation between the
caustic in the stellar material and the second caustic in the dark
matter. Firstly, we must consider that the stars and dark matter do
not undergo the same evolution in the build-up of a halo. A
significant amount of the dark matter is assembled by ``smooth''
accretion \citep[e.g.][]{angulo10,genel10, wang11}. On the other hand,
the stars are assembled from a ``lumpier'' accretion process, as the
stars do not populate subhaloes below a certain mass threshold
\citep[e.g.][]{sawala15}. Secondly, the stripping of stars from bound
subhaloes proceeds differently to the stripping of the dark matter
\citep[e.g.][]{penarrubia08,fattahi18}: the (less bound) dark matter
is stripped first, and the more centrally concentrated stars are
almost always stripped close to pericentre, when almost all of the
dark matter has already been peeled away. 

We speculate that to lose stars to tidal forces subhaloes must
typically pass through at least two pericentres, and thus the ``edge''
of the stellar material coincides with the second caustic in the dark
matter. This may be especially true in relatively major mergers, which
typically dominate the mass budget of the accreted stellar halo (see
e.g. \citealt{purcell07, cooper10, deason16, dsouza18}), when such passages lead
to a loss of angular momentum and shrinking of the
pericentre. Finally, we remark that the relation between the ``edges''
of stars and dark matter may vary at different mass scales. Here, we
have focused on Milky Way-mass halos, but the non-linear stellar
mass to dark matter mass relation \citep{behroozi10, moster10,
  read17}, and the varying smooth to lumpy mass accretion rate
\citep{genel10}, will likely lead to different relations at higher and
lower masses.

In Section~\ref{sec:dm} we discussed how the second caustic of the
dark matter, which we now see coincides with the stellar caustic, may be
the most relevant definition of the edge of the Milky Way. This means
that the edge of our own Galaxy is, potentially, observable in the
stellar distribution. Currently, the density profile of the stellar
halo has only been mapped out to $\sim \! \! 50-100$ kpc
\citep[e.g.][]{deason11, sesar11, deason14, xue15, slater16,
  hernitschek18}. Moreover, radial velocities of stars are only
available, in any significant numbers, out to similar distances
\citep[e.g.][]{mauron04, deason12, bochanski14, cohen17}. However,
with upcoming wide-field photometric and spectroscopic facilities like the Rubin Observatory Legacy Survey of Space and Time \citep[LSST,][]{lsst}, the Roman Space Telescope \citep{wfirst}, the Mauna Kea
Spectroscopic Explorer \citep[MSE,][]{mse} and the Subaru Prime Focus
Spectrograph \citep[PFS,][]{pfs} on the horizon, exploring these
extreme distances will be feasible in the near future.

Finally, it is worth discussing how the concept of galaxy edge is
relevant to studies that require a definition of where the halo
ends. For example, when using the escape velocity of local halo stars
to estimate the total mass of the Galaxy, the definition of the radius
of ``escape'' is an important element of the analysis. Indeed,
\cite{deason19} used a radius of $2r_{\rm 200c}$
($\sim \! \! 1.25 r_{\rm 200m}$), which is at the extreme end for the
Auriga haloes. However, while this approach is conservative in that it
does not allow for radii where stars can potentially escape, our
results suggest that a smaller radius is likely more applicable. For
example, the median stellar caustic radius of the Auriga simulations
is $0.7r_{\rm 200m}$, which is approximately $1.2r_{\rm 200c}$. If
this distance is used in the \cite{deason19} analysis to define the
radius beyond which stars have escaped, then the total mass of the
Milky Way is revised upwards by 20 percent. Interestingly, this is
approximately the change that \cite{grand19} found was required to
correct the mass estimates when the procedure is applied to the Auriga
haloes. In particular, \cite{grand19} suggest that the mass estimates
are underestimated because the local stars do not reach out to
$2r_{\rm 200c}$. Here, we show that this is indeed the case. However,
as a cautionary note, we should use the observed
$R^{\rm STAR}_{\rm Caustic}$ rather than the median value of the
Auriga haloes, which does not necessarily coincide with the true Milky
Way value (see end of Section \ref{sec:subs}). Finally, we note that the term ``escape velocity'' is a misleading term when discussing the highest velocity halo stars. In reality, much faster stars would not ``escape'' as such, but rather they just do not exist in the stellar distribution.

\subsection{Subhaloes and Dwarf Galaxies}
\label{sec:subs}
\begin{figure}
  \centering
        \includegraphics[width=\linewidth,angle=0]{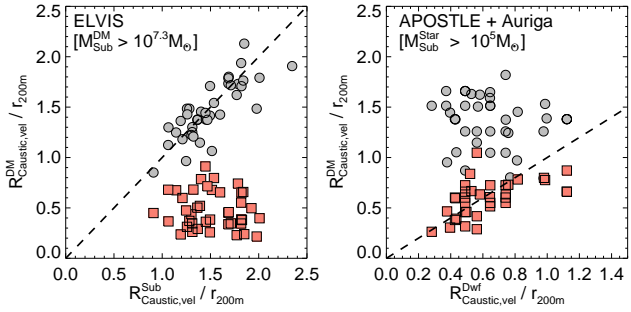}
        \caption[]{The radius of the dark matter caustics against that
          of subhalo/dwarf caustics. We focus only on the velocity
          caustics which are more easily identified with low numbers
          of tracers. In the left-hand panel, we show all subhaloes in
          ELVIS down to the resolution limit.  Dwarfs, defined as
          subhaloes with at least one star particle, in the Auriga and
          APOSTLE simulations are shown in the right-hand panel. We
          only aim to identity one caustic for the subhalo
          populations, owing to low-number statistics. Thus, the two
          types of symbols shown in these plots relate to the two
          caustics in the dark matter (splashback = filled grey
          circles, second caustic = filled red squares). For the
          subhaloes in ELVIS (left-panel), this generally corresponds
          to the splashback radius (filled grey circles) of the dark
          matter particles. However, the caustic defined by the
          luminous dwarfs in APOSTLE and Auriga (right panel)
          corresponds to the second caustic (filled red squares) in
          the dark matter. Note in several cases ($\sim \! \! 30$
          percent) a caustic could not be identified in the luminous
          dwarfs, most commonly due to low numbers.}
          \label{fig:subs_caustics}
\end{figure}

\begin{figure}
    \begin{minipage}{\linewidth}
        \centering
        \includegraphics[width=\textwidth,angle=0]{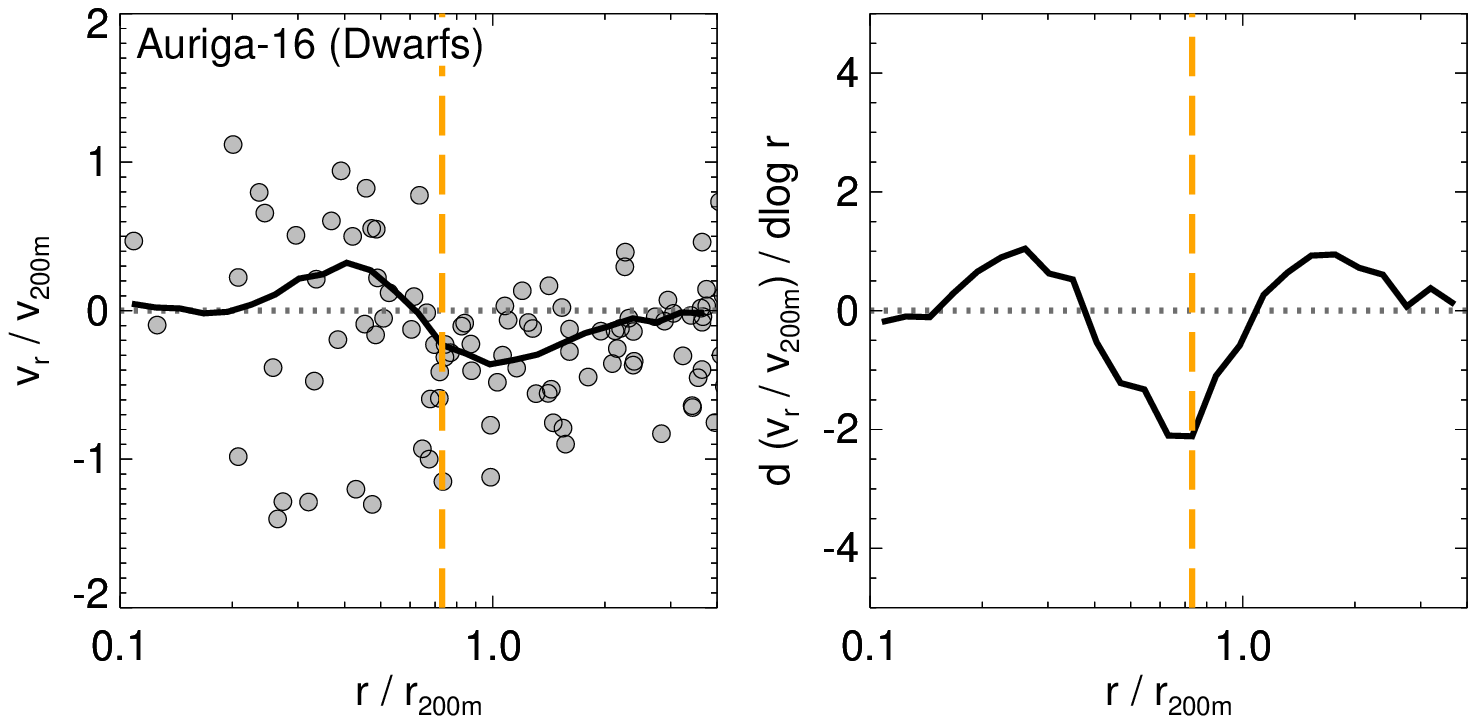}
    \end{minipage}
    \begin{minipage}{\linewidth}
        \centering
        \includegraphics[width=\textwidth,angle=0]{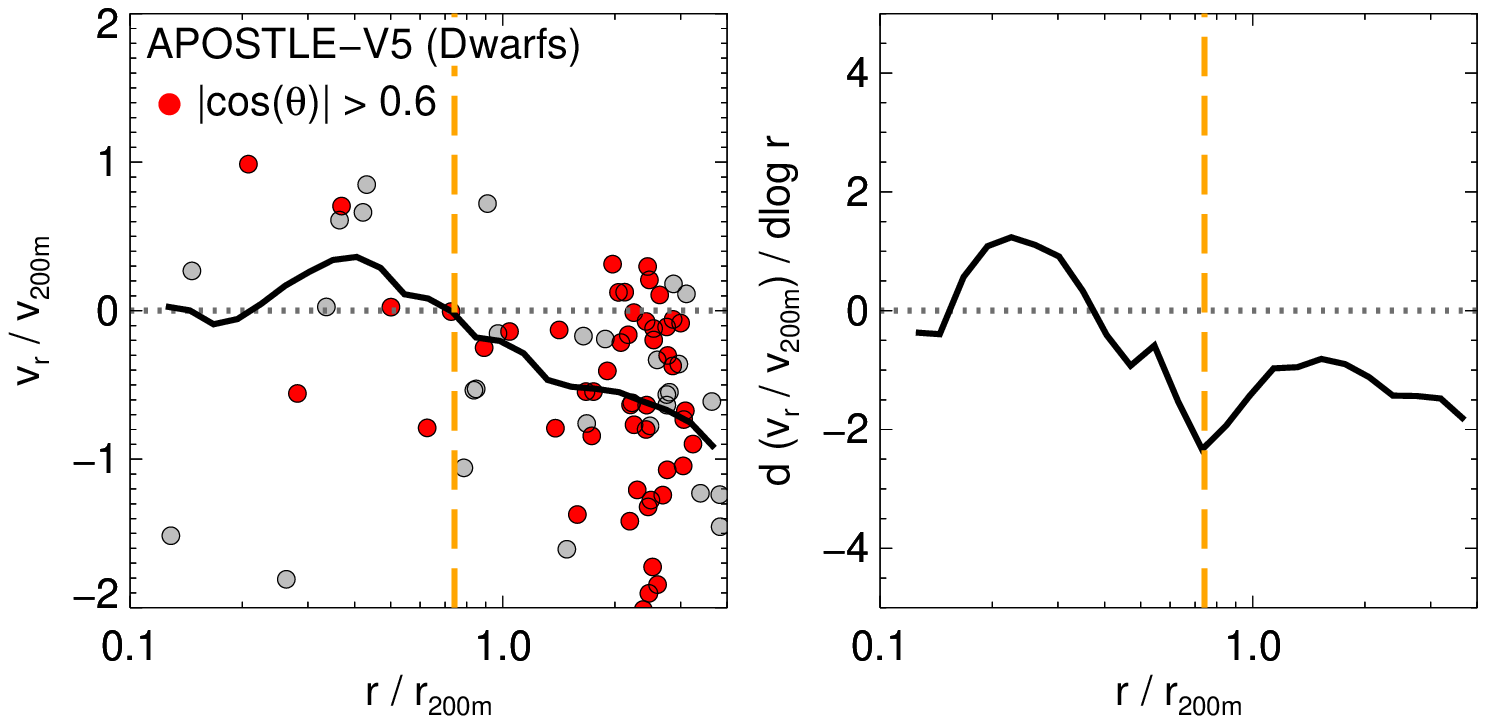}
    \end{minipage}
        \caption[]{Examples of caustics defined from the luminous
          dwarf population in Auriga (Au-16, top) and APOSTLE (V5,
          bottom). The left panels show the radial velocities of the
          dwarfs as a function of radius. For the paired halo dwarfs
          with $|\mathrm{cos}(\theta) > 0.6|$ (i.e close in angle to
          the line joining the two haloes) are indicated in red. The
          right-hand panels show the logarithmic slope profiles of the
          radial velocities. The vertical dashed line indicates the
          caustic.} 
        \label{fig:subs_eg}
\end{figure}

In the previous subsections, we have focused on the distribution of
dark matter and stars. Now we apply a similar analysis to the subhalo
population. In this case, the number of discrete tracers is much lower
than for the dark matter or star particles. For this reason, we
concentrate only on the caustics defined in velocity space, where it
is easier to identify features associated with caustics when there are
low numbers of tracers. It is worth noting that there is no division
into position angle sectors here (cf. the dark matter and stars),
which makes the subhalo-based profiles sensitive to
substructure. Thus, although this analysis is a valuable first step,
we plan to apply more sophisticated techniques tailored towards highly
discretely sampled distributions in future work.

We use the (dark matter only) ELVIS suite to study the general subhalo
population, and APOSTLE and Auriga to analyse the ``dwarf''
population. Here we do not distinguish between isolated and paired
environments and, in the paired cases, only consider subhaloes with
$|\mathrm{cos}(\theta)| < 0.6$ to identify caustics. We define
subhaloes as all bound substructures with
$M^{\rm DM}_{\rm Sub} > 10^{7.3}\mathrm{M}_\odot$. This is the
convergence limit for subhaloes found by \cite{gk14}. In APOSTLE and
Auriga, subhaloes with at least one star particle are identified as
luminous dwarfs. This approximately corresponds to subhaloes with
$M^{\rm Star}_{\rm Sub} > 10^{5}\mathrm{M}_\odot$.

For each individual halo we use the logarithmic slope of the radial
velocity profile to define the caustics in the subhalo
population. Note that for the dwarf galaxies, where the numbers of
objects are typically low (O(100) per halo), we change the binning in
logarithmic radius to have 25 equally spaced bins in the range
$\mathrm{log}(r/r_{\rm 200m}) \in [-1.0, 0.5]$ and use the same
smoothing kernel as in the previous subsections. Due to the small
numbers we only identify the most prominent caustic and do not attempt
to find two distinct caustics.

The resulting caustics are shown in Fig.~\ref{fig:subs_caustics} as a
function of the (two) dark matter caustics (computed in Section
\ref{sec:dm}). Caustics can be identified for the majority of subhalo
populations, but in several cases (30 percent) a caustic could not be
identified in the dwarf population, mainly as a result of small
numbers. The filled gray circles in Fig.~\ref{fig:subs_caustics}
indicate the splashback radii in the dark matter and the filled red
squares the second caustic in the dark matter. Interestingly, we find
that the caustic in the subhalo population corresponds to the
splashback radius (left panel), while the caustic in the luminous
dwarfs' population coincides with the second caustic in the dark
matter (right panel). This is perhaps unsurprising as the subhalo
population traces the dark matter, while the luminous dwarfs are more
closely related to the accretion of the more massive subhaloes, and
hence the stellar halo.

\begin{figure}
  \centering
  \includegraphics[width=\linewidth,angle=0]{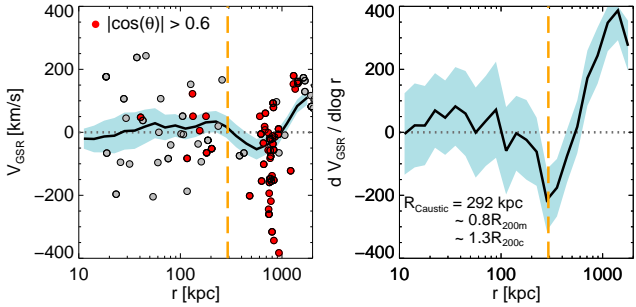}
  \caption[]{\textit{Left panel:} the radial velocities (in Galacocentric coordinates, $V_{\rm GSR}$) of observed
    Local Group dwarf galaxies. Dwarfs with
    $|\mathrm{cos}(\theta) > 0.6|$ (i.e close in angle to the line
    joining the Milky Way and M31) are indicated in red. The solid
    black line indicates the median radial velocity
    profile, and the shaded region indicates the dispersion (defined as $1.4826 \times$ the median absolute deviation) calculating using a bootstrap method. \textit{Right panel:} the logarithmic slope profile of
    the median radial velocity. The vertical dashed line indicates the
    caustic that defines the edge of the Galaxy. This lies at
    $290$~kpc and approximately corresponds to $0.8r_{\rm 200m}$ (or
    $\sim \! \! 1.0 r_{\rm vir}$, $\sim \! \! 1.3r_{\rm 200c}$),
    assuming the Milky Way mass estimated by  \cite{callingham19}. }
          \label{fig:obs}
\end{figure}

We show two examples for the dwarf galaxy population in Auriga-16 (top
panel) and APOSTLE-V5 (bottom panel; centred centred on
$(x,y,z)=(42.867, 88.474, 93.675)$~Mpc) in Fig.~\ref{fig:subs_eg}. The
caustics are not as clearly defined as in the dark matter or stars,
but, importantly, there are already observations of luminous dwarf
tracers out to large distances in the Local Group, so this analysis is
observationally motivated. In Fig. ~\ref{fig:obs} we perform the same
analysis on the \textit{observed} dwarfs. Here, we use the latest
compilation of dwarfs from \cite{mcconnachie12}, and show physical
radius and velocity (rather than in units of $r_{\rm 200m}$ and
$v_{\rm 200m}$).  The distances and and radial velocities are converted to Galactocentric coordinates, assuming a circular velocity of $v_c(r_0)= 235$ km s$^{-1}$ at the position of the Sun ($r_0=8.1$ kpc), and a peculiar solar motion of ($U_\odot, V_\odot, W_\odot) = (11.1, 12.24, 7.25)$ km s$^{-1}$ \citep{schonrich10}. We use 22 radial bins equally spaced in
$\mathrm{log} (r)$ between $1.0$ and $3.3$.  As we did previously, the
logarithmic slope profile is computed using the fourth-order
Savitzky–Golay smoothing algorithm over the 15 nearest bins
\citep{savitzky64}. We indicate in the figure dwarfs which are close
in angle to the line joining the Milky Way and M31 (i.e
$|\mathrm{cos}(\theta) > 0.6|$). In practice, we find little
difference if we include or exclude these dwarfs. 

We identify a minimum in the observed population of dwarfs at
$\sim \! \! 290$~kpc. Using a bootstrap method to estimate the
uncertainty, we find $R_{\rm edge} = 292 \pm 61$~kpc. If we assume the
Milky Way halo mass recently measured by \cite{callingham19} and a
typical halo concentration ($\sim \! \! 10$ for Milky Way-mass haloes;
\citealt[e.g.][]{neto07,ludlow14, klypin16}), this radius corresponds
to $0.8r_{\rm 200m}$ (or $1.3r_{\rm 200c}$). Interestingly, this
radius (292 kpc) lies at exactly the ``virial radius'' defined by the
fitting formulae in \cite{bryan98}. Moreover, this also coincides with the radius where the H\textsc{i} content of Local Group satellites sharply drops (around 270 kpc, \citealt{grcevich09}). Given the rather large uncertainty in the measurement, these could simply be coincidences, however, it is worth noting that we are probing an interesting radial regime of the Galactic halo. 

We can also use this measured radius to independently estimate the mass of the Milky Way using the escape velocity analysis described by \cite{deason19}. As mentioned in Section \ref{sec:stars}, this technique depends on the definition of the ``outer boundary'' of the halo stars. If we use a boundary of 290 kpc, rather than a fixed fraction of $r_{\rm 200c}$ like \cite{deason19}, we find a mass of $M_{\rm 200c} \sim 1.1 \times 10^{12}M_\odot$. Although there is considerable uncertainty in this definition of halo edge, it is reassuring that this mass is in excellent agreement with the recent measurements by \cite{callingham19} and \cite{cautun20}.

While we suggest that the edge of the Milky Way halo lies at 290 kpc,
this remains a tentative result for two important reasons. Firstly,
the value is strongly dependent on Leo I (located at 250 kpc): there
is a significant gap between the most distant satellite of the Milky
Way and the nearest dwarfs in the Local Group. Secondly, and perhaps
most importantly, our census of local dwarfs is far from complete and
we have made no attempt to correct for selection effects or
observational biases. Indeed, as recently predicted by
\cite{fattahi20}, there are troves of local group dwarfs waiting to
be discovered by future wide-field imaging surveys.

\section{Conclusions}
\label{sec:conc}
In this work we have analysed three different suites of simulated
Milky Way-mass haloes (ELVIS, APOSTLE and Auriga) to explore the
``edge'' of Galactic-sized haloes. We use the logarithmic slope
profiles of the density and radial velocity distributions to
identify the location of caustics in the halo. These features,
which correspond to the build up of particles at apocentre, are used
to define the edges of the dark matter, stars, and subhalo
population. Our main conclusions are summarised as follows:

\begin{itemize}
\item We typically identify two distinct caustics in the outer dark
  matter profiles. The outermost caustic, called the ``splashback''
  radius, is the boundary at which accreted dark matter reaches its
  first orbital apocentre after turnaround. This lies at approximately
  $\sim \! \! 1.4r_{\rm 200m}$ for Milky Way-mass haloes. We suggest that the second
  caustic, which is located at a smaller radius
  ($\sim \! \! 0.6r_{\rm 200m} \approx r_{\rm 200c}$) and is typically
  less prominent than the caustic at the splashback radius,
  corresponds to the edge of the material which has passed through at
  least two pericentric passages. 
\item In Local Group-like environments, the splashback radius of one
  of the haloes is poorly defined, as it often overlaps with the other
  halo.  However, the second caustic in the dark matter is less affected 
  by the companion and appears to be a more useful choice for the
  definition of the halo boundary of the Milky Way.
\item We identify a prominent caustic in the stellar distribution in
  both the radial density and velocity profiles. This typically lies
  at $0.6r_{\rm 200m}$ and, in the majority of cases, coincides with
  the second caustic of the dark matter. This feature can potentially
  be identified in the Milky Way using future observational
  facilities, such as LSST and MSE. Moreover, there
  is scope to measure this edge in external galaxies, either by
  stacking profiles, or by obtaining deeper and wider images with
  forthcoming facilities such as the Roman Space Telescope.
\item The outer caustic, corresponding to the splashback radius, can
  be identified in the phase-space distribution of the subhalo
  population. If we consider only luminous dwarfs (with
  $M_{\rm star} > 10^{5} \mathrm{M}_\odot$) the best defined caustic
  coincides with the second caustic in the dark matter (and hence with
  the stellar caustic).
\item We applied our analysis to the currently known population of
  dwarf galaxies in the Local Group. We predict that the edge of the
  Milky Way (defined as the second caustic in the dark matter) lies at
  $\sim \! \! 290$ kpc. For the total Milky Way mass measurement by
  \cite{callingham19}, this radius coincides approximately with the
  value of $r_{\rm vir}$ obtained from the fitting formula of
  \cite{bryan98}, albeit with significant uncertainty. This is a
  tentative measurement of the Galactic edge, but will greatly improve
  with future discoveries of more Local Group dwarfs.

\end{itemize}

In many analyses of the Milky Way halo its outer boundary is a
fundamental constraint. Often the choice is subjective, but as we have
argued, it is preferable to define a physically and/or observationally
motivated outer edge. Here we have linked the boundary of the
underlying dark matter distribution to the observable stellar halo and
the dwarf galaxy population.  There is great hope that future data
will provide a more robust and accurate \textit{measurement} of the
edge of the Milky Way and nearby Milky Way-mass galaxies than the one
we have presented here.  In this work we have focused on Milky Way
mass haloes in a $\Lambda$CDM cosmology, but a similar analysis can be
extended to wider mass scales and applied to different cosmologies or
dark matter models.

\section*{Acknowledgements}
AD thanks Andrey Kravtsov for many enlightening science discussions on this topic, and Phil Mansfield for his valuable input.  We thank an anonymous referee for improving the clarity of this paper.

AD is supported by a Royal Society University Research Fellowship, and
AD, AF, CSF and KO by the Science and Technology Facilities Council (STFC)
[grant numbers ST/F001166/1, ST/I00162X/1,ST/P000541/1]. CSF and KO are also
supported by ERC Advanced Investigator grant, DMIDAS [GA 786910]. 
This work used the DiRAC Data Centric system at Durham University,
operated by the ICC on behalf of the STFC DiRAC HPC Facility
(www.dirac.ac.uk). This equipment was funded by BIS National
E-infrastructure capital grant ST/K00042X/1, STFC capital grant
ST/H008519/1, and STFC DiRAC Operations grant ST/K003267/1 and Durham
University. DiRAC is part of the National E-Infrastructure. 

AD thanks
the staff at the Durham University Day Nursery who play a key role in
enabling research like this to happen.

\section*{Data Availability Statement}
The data presented in the figures are available upon request from the corresponding author. The raw simulation data can be requested from the ELVIS \citep{gk14},  APOSTLE \citep{fattahi16, sawala16}, and Auriga \citep{grand17} teams.

\bibliographystyle{mnras}
\bibliography{mybib}

\label{lastpage}
\end{document}